\documentclass[twocolumn, tighten]{aastex61}

\usepackage{bm}

\received{2018 March 9}
\revised{2018 April 26}
\accepted{2018 May 6}
\published{2018 June 14}

\shorttitle{Opacity of fractal dust aggregates}
\shortauthors{Tazaki and Tanaka}

\begin{document}

\title{Light scattering by fractal dust aggregates. II. Opacity and asymmetry parameter}

\correspondingauthor{Ryo Tazaki}
\email{rtazaki@astr.tohoku.ac.jp}

\author[0000-0003-1451-6836]{Ryo Tazaki}
\affil{Astronomical Institute, Graduate School of Science
Tohoku University, 6-3 Aramaki, Aoba-ku, Sendai 980-8578, Japan}

\author{Hidekazu Tanaka}
\affil{Astronomical Institute, Graduate School of Science
Tohoku University, 6-3 Aramaki, Aoba-ku, Sendai 980-8578, Japan}

\nocollaboration



\begin{abstract}
Optical properties of dust aggregates are important at various astrophysical environments. 
To find a reliable approximation method for optical properties of dust aggregates,
we calculate the opacity and the asymmetry parameter of dust aggregates by using a rigorous numerical method, the T-Matrix Method (TMM), and then the results are compared to those obtained by approximate methods; the Rayleigh--Gans--Debye (RGD) theory, the effective medium theory (EMT), and the distribution of hollow spheres method (DHS).
First of all, we confirm that the RGD theory breaks down when multiple scattering is important. In addition, we find that both EMT and DHS fail to reproduce the optical properties of dust aggregates with fractal dimension of 2 when the incident wavelength is shorter than the aggregate radius.
In order to solve these problems, we test the mean field theory (MFT), where multiple scattering can be taken into account.
We show that the extinction opacity of dust aggregates can be well reproduced by MFT. 
However, it is also shown that MFT is not able to reproduce the scattering and absorption opacities when multiple scattering is important.
We successfully resolve this weak point of MFT, by newly developing a modified mean field theory (MMF).
Hence, we conclude that MMF can be a useful tool to investigate radiative transfer properties of various astrophysical environments. 
We also point out an enhancement of the absorption opacity of dust aggregates in the Rayleigh domain, which would be important to explain the large millimeter-wave opacity inferred from observations of protoplanetary disks.
\end{abstract}




\section{Introduction}
Dust aggregates are ubiquitous at various astrophysical environments where coagulation of dust particles can take place, such as molecular clouds \citep{Ormel:2009aa, Ormel:2011aa, Steinacker:2010aa}, protoplanetary disks \citep{Blum:2008aa, Okuzumi:2012aa}, and planetary atmospheres \citep{Marley:2013aa}. Optical properties of dust aggregates are key ingredients of the radiative transfer in these environments, and these often govern observational appearance of the environments. 

Optical properties of dust aggregates have been calculated by using numerical methods, such as the T-Matrix Method \citep[TMM; ][]{mishchenko96} and the Discrete Dipole Approximation \citep[DDA; ][]{purcell73, draine94}. 
By using these techniques, a multitude of numerical studies have been performed \citep[e.g.,][]{Kimura:2016aa, Silsbee:2016aa}. 
Unfortunately, these numerical methods are still computationally demanding, in particular, when the size parameter of aggregate particles becomes large, and hence approximate methods are useful for many astrophysical purposes. 
Many approximate methods have been proposed so far \citep[e.g.,][]{B83, Ossenkopf91, Mukai:1992aa, Stognienko:1995aa, B97, vosh05, vosh07, M03, M05, Min:2008aa, M16}; however, it still remains unclear which of them can reliably reproduce optical properties of aggregate particles.

In the series of this paper, we study optical properties of dust aggregates using TMM and investigate a fast and reliable approximate method which is able to reproduce the TMM results. A byproduct of finding a reliable approximation is that it leads to better, or sometimes intuitive, understanding of their optical properties. Tazaki et al. \citep[2016, henceforth][]{T16} studied the angular dependence of the scattering of fractal dust aggregates, such as the phase function and the degree of polarization and showed that the Rayleigh--Gans--Debye (RGD) theory is an useful method for calculating these properties. 
This study examines the angle-integrated optical properties of dust aggregates, such as the opacity and the asymmetry parameter.
 
This paper also attempts to relax a limitation of the RGD theory studied in \citet{T16}.
Because the RGD theory is a single scattering theory, it is applicable when multiple scattering is sub-dominant.
This limitation is expected to be relaxed by considering the multiple scattering effect. 
For this purpose, we adopt the mean field assumption \citep{B86}. With this assumption, multiple scattering can be solved self-consistently, while keeping most of the formulation of the RGD theory preserved \citep{B97}. We refer to this approach as the mean field theory (MFT). In this paper, we test the validity of MFT and discuss whether it is a reliable approximation or not. If not, we propose an alternative method.

Furthermore, we test the validity of other widely used approximate methods, such as the effective medium theory \citep[EMT; ][]{Mukai:1992aa, Henning96, Min:2008aa} and the distribution of hollow spheres method \citep[DHS; ][]{M03, M05, M16}. For example, these methods have been commonly used in both observational and theoretical studies of circumstellar environments \citep[e.g..][]{Ormel:2011aa, Min:2012aa, K14, cuzzi14, Woitke:2016aa}.

In the EMT method, a dust aggregate is approximated by a single homogeneous sphere, and then the Mie theory \citep{B83} is applied to obtain the optical properties. In this method, information of the dust aggregate, such as the size, the porosity and the composition, is reduced to a quantity, so-called the effective dielectric function.
\citet{T16} showed that EMT cannot reproduce the phase function of fluffy dust aggregates. 
In this paper, we study the validity of EMT in the calculations of the opacity and the asymmetry parameter.
Our primary focus is large relatively porous dust aggregates; ballistic cluster cluster agglomerates (BCCA) and ballistic particle cluster agglomerates (BPCA). \citet{shen08} studied the validity of EMT for small and compact ballistic dust aggregates, such as BPCA and more compact dust aggregates.

DHS is an approximate method to mimic optical properties of irregularly shaped particles \citep{M05, M16}. 
In the DHS method, optical properties of dust aggregates are obtained by using those of hollow spheres.
DHS seems to capture optical properties of irregularly shaped particles in the Rayleigh domain, such as an enhancement of the absorption opacity and the red shifted solid state feature \citep{M03, Min:2006aa}.
 The applicability of DHS to the larger particles has been mainly tested using compact dust aggregates (porosity of 25\% in \citet{M16}); therefore, in this paper, we test the applicability of DHS to large fluffy dust aggregates (porosity $\gtrsim85\%$).

It is known that coagulation of dust particles leads to an enhancement of the absorption opacity in the Rayleigh domain \citep{Wright:1987aa,bazell90,kozasa92, Mackowski:1995aa, Henning:1995aa, Stognienko:1995aa, Mackowski:2006aa, Kohler:2011aa}. 
In the case of protoplanetary disks, a value of the absorption opacity at millimeter wavelength is important because it is often used to derive the disk dust mass \citep{Beckwith:1990aa}. We investigate how an enhancement of the absorption opacity of the dust aggregate depends on its number of monomers, structure, and composition. In addition, we also discuss the methods that might be used to approximate this effect.

This paper is organized as follows. 
In Section \ref{sec:models}, we summarize light scattering models of fractal dust aggregates based on the RGD theory and MFT.
In Section \ref{sec:opacity}, we show that MFT can reproduce the extinction opacity; however, MFT fails to reproduce the scattering and absorption opacities. 
We propose a modification to MFT, a modified mean field theory (MMF), whose validity is also tested in this section. 
In Section \ref{sec:approx}, we compare the optical properties obtained by other approximate methods (EMT and DHS) to those obtained by TMM and discuss their validity.
In Section \ref{sec:excess}, we study an enhancement of the absorption opacity of dust aggregates in the Rayleigh domain.
Section \ref{sec:conclusion} summarizes the results of this study.
Table \ref{tab:namelist} shows the list of abbreviations of various methods used in this paper.

\begin{table*}[htbp]
\renewcommand{\thetable}{\arabic{table}}
\centering
\caption{Computational methods used in this paper} \label{tab:namelist}
\begin{tabular}{ccc}
\tablewidth{0pt}
\hline
\hline
Acronym & Name & Relevant paper(s)\\ 
\hline
\decimals
TMM & T-Matrix Method  & \citet{M96}  \\
RGD &  Rayleigh--Gans--Debye theory  & \citet{T16} \\
MFT &  Mean Field Theory & \citet{B86, B97} \\
MMF &  Modified Mean Field Theory & This study \\
EMT (MG-Mie) &  Mie theory with Maxwell-Garnett mixing rule & \citet{Mukai:1992aa, K14} \\
EMT (APMR-Mie) &  Mie theory with aggregate polarizability mixing rule & \citet{Min:2008aa} \\
DHS &  Distribution of hollow spheres & \citet{M05, M16} \\
CDE &  Continuous distribution of ellipsoids & \citet{B83} \\ 
\hline
\end{tabular}
\end{table*}

\section{Light scattering models} \label{sec:models}
Light scattering models for fractal dust aggregates are presented. For the sake of simplicity, we consider dust aggregates of identical monomers.
For many astrophysical purposes, orientation averaged optical properties of dust aggregates are important. Hence, we consider optical properties of the dust aggregates averaged over random orientations.

Light scattering models described below are based on {\it the statistical distribution of monomers} in the aggregate. Once the optical properties are averaged over random orientations, these properties are well described by the aggregate model with a statistically isotropic distribution of monomers. As a consequence, the random orientation assumption makes light scattering models rather simple, e.g., scattering properties do not depend on the zenith angle \citep[see e.g.,][]{T16}. A caveat of the statistically isotropic assumption is that applications to dichroic extinction or polarized emission are precluded.

The calculation of MFT has the following three steps.
The first step is to calculate optical properties of a monomer particle (Section \ref{sec:mie}).
At the second step, interference of scattered waves from each monomer is calculated under the single scattering assumption (Section \ref{sec:rgd}), where the statistical arrangement of monomers is assumed (see Section \ref{sec:gu}). The final step is to calculate multiple scattering using the mean field approximation (Section \ref{sec:mft}). Note that the first and second steps correspond to the RGD theory \citep{T16}. 

\subsection{Optical properties of a spherical monomer} \label{sec:mie}
Light scattering models presented below are applicable to a single monomer size and single isotropic composition.
In this paper, all monomers are assumed to be spherical, and hence the Mie theory \citep{B83} can be used to obtain optical properties of the monomer.

In what follows, we adopt the same notation as \citet{B83}. 
For an isolated sphere, the expansion coefficients of the scattered field in vector spherical harmonics are related to those of the incident field via the Lorenz--Mie coefficients ($a_n, b_n$) \citep{B83}:
\begin{eqnarray}
a_n&=&\frac{m\psi_n(mx)\psi_n'(x)-\psi_n(x)\psi'_n(mx)}{m\psi_n(mx)\xi_n'(x)-\xi_n(x)\psi'_n(mx)},\label{eq:an} \\ 
b_n&=&\frac{\psi_n(mx)\psi_n'(x)-m\psi_n(x)\psi'_n(mx)}{\psi_n(mx)\xi_n'(x)-m\xi_n(x)\psi'_n(mx)}, \label{eq:bn}
\end{eqnarray}
where $m$ is the complex refractive index, $x=kR_0$ is the size parameter of the monomer, $R_0$ is the radius of the monomer, $k$ is the wavenumber in vacuum, and $\psi$ and $\xi$ are the Riccati--Bessel functions. Once the Lorenz--Mie coefficients are obtained, optical properties of the monomer, e.g., the opacity and the scattering matrix, are immediately derived \citep[e.g.,][]{B83}.
For example, the extinction cross section can be obtained by the optical theorem:
\begin{equation}
C_{\mathrm{ext}}^0=\frac{2\pi}{k^2}\sum_{n=1}^{\infty}(2n+1)\mathrm{Re}(a_{n}+b_{n}) \label{eq:Cext},
\end{equation}
where $C_{\mathrm{ext}}^0$ represents the extinction cross section of the individual monomer and superscript ``0" denotes a single monomer.
The elements of the amplitude scattering matrix are
\begin{eqnarray}
S_1^0&=&\sum_{n=1}^{\infty}\frac{2n+1}{n(n+1)}(a_{n}\pi_n+b_{n}\tau_n), \label{eq:S1}\\
S_2^0&=&\sum_{n=1}^{\infty}\frac{2n+1}{n(n+1)}(a_{n}\tau_n+b_{n}\pi_n), \label{eq:S2}
\end{eqnarray}
where $\pi_n=P_n^1/\sin\theta$ and $\tau=dP_n^1/d\theta$, where $P_n^1$ is the Legendre Polynomials. 
Using Equations (\ref{eq:S1} and \ref{eq:S2}), one can calculate scattering matrix elements \citep[see Chapter 3 of][]{B83}.
For example, the (1,1) element of the scattering matrix, $S_{11}$, which is related to the scattered intensity, can be expressed as
\begin{equation}
S_{11}^0=\frac{1}{2}(|S_1^0|^2+|S_2^0|^2) \label{eq:monos11}.
\end{equation}

\subsection{RGD theory} \label{sec:rgd}
The RGD theory, or the first order of the Born approximation, calculates the interference of scattered waves from every monomer taking the aggregate structure into account \citep[e.g.,][]{T16}.
In the RGD theory, the aggregate structure is specified by using the two-points correlation function.
The RGD theory can be applicable when \footnote{The criterion was slightly modified from \citet{T16} based on the argument presented in Sections \ref{sec:wavel} and \ref{sec:syserror}.}
\begin{eqnarray} 
|m-1|&<&2, \label{eq:rgdcond1}\\
\Delta\phi&<&1,\label{eq:rgdcond2}
\end{eqnarray}
where $\Delta\phi$ is the (maximum) phase shift of the dust aggregate defined by
\begin{equation}
\Delta\phi=\max(2x|m-1|, 2x_c|m_{\rm eff}-1|), \label{eq:phase}
\end{equation}
where $x_c=kR_c$ is the size parameter of the dust aggregate, $R_c$ is the characteristic radius of the dust aggregate defined by $R_c=\sqrt{5/3}R_g$, where $R_g$ is the radius of gyration defined by Equation (14) in \citet{T16}, and $m_{\rm eff}$ is the effective refractive index of the dust aggregate calculated by using the Maxwell-Garnett mixing rule \citep{MG09}. 

In the RGD theory, the absorption cross section of the dust aggregate, $C_{\mathrm{abs}}^{(\mathrm{RGD})}$, can be written as
\begin{equation}
C_{\mathrm{abs}}^{(\mathrm{RGD})}=NC_{\mathrm{abs}}^0 \label{eq:absrgd}
\end{equation}
where $N$ is the number of monomers and $C_{\mathrm{abs}}^0$ represents the absorption cross section of the individual monomer \citep{B86}.
It follows from Equation (\ref{eq:absrgd}) that the mass absorption opacity of dust aggregates becomes exactly the same as that of the monomer.
The scattering cross section of the dust aggregate, $C_{\mathrm{sca}}$, is defined as
\begin{equation} 
C_{\rm sca}=\frac{2\pi}{k^2}\int_0^\pi S_{11, {\rm agg}}(\theta)\sin\theta d\theta \label{eq:scargd}
\end{equation}
where $\theta$ is the scattering angle and $S_{11, {\rm agg}}$ is the (1,1) element of the scattering matrix of dust aggregates.
As shown in \citet{T16}, the (1,1) element of the scattering matrix of dust aggregates obtained by the RGD theory,  
$S_{11, {\rm agg}}^{(\mathrm{RGD})}$, can be written by
\begin{equation}
S_{11, {\rm agg}}^{(\mathrm{RGD})}(\theta)=N^2S_{11}^0(\theta)\mathcal{S}(q), \label{eq:RGDtheory}
\end{equation}
where $\mathcal{S}(q)$ is the static structure factor (or the simply structure factor) and $q=2k\sin(\theta/2)$ is a magnitude of the scattering vector. The static structure factor specifies the statistical arrangement of monomers in the aggregate, and it is equivalent to the Fourier transform of the two-points correlation function (see Section \ref{sec:gu} for more detail). The scattering cross section of the dust aggregate in the RGD theory, $C_{\mathrm{sca}}^{(\mathrm{RGD})}$, can be obtained by substituting Equation (\ref{eq:RGDtheory}) into Equation (\ref{eq:scargd}).
The extinction cross section is then calculated by $C_{\rm ext}^{(\mathrm{RGD})}=C_{\rm sca}^{(\mathrm{RGD})}+C_{\rm abs}^{(\mathrm{RGD)}}$.

\subsection{MFT} \label{sec:mft}
The effect of multiple scattering can be readily implemented by means of the mean field approximation  \citep{B86, B97}.
One advantage of the mean field approximation is that it preserves the optical theorem.
Conversion from the RGD theory to MFT can be achieved by replacing the Lorenz--Mie coefficients of the monomer by
\begin{equation}
(a_n,b_n) \rightarrow (\bar{d}_{1,n}^{(1)}, \bar{d}_{1,n}^{(2)}) \label{eq:rgdmft}
\end{equation}
where $\bar{d}_{1,n}^{(1)}$ and $\bar{d}_{1,n}^{(2)}$ are coefficients of the mean field obtained from a set of linear equations \citep{B97}
\begin{eqnarray}
\bar{d}_{1,n}^{(1)}&=&a_n\left\{1-(N-1)\sum_{\nu=1}^{\infty} \bar{A}_{1,n}^{1,\nu}\bar{d}_{1,\nu}^{(1)}+\bar{B}_{1,n}^{1,\nu}\bar{d}_{1,\nu}^{(2)}\right\} \label{eq:basic1},\\
\bar{d}_{1,n}^{(2)}&=&b_n\left\{1-(N-1)\sum_{\nu=1}^{\infty} \bar{B}_{1,n}^{1,\nu}\bar{d}_{1,\nu}^{(1)}+\bar{A}_{1,n}^{1,\nu}\bar{d}_{1,\nu}^{(2)}\right\} \label{eq:basic2},
\end{eqnarray}
where coefficients $\bar{A}_{1,n}^{1,\nu}$ and $\bar{B}_{1,n}^{1,\nu}$ are defined in Appendix \ref{sec:app1}. 
The first terms in the parenthesis of right hand side of Equations (\ref{eq:basic1}, \ref{eq:basic2}) correspond to the external field, and the second terms represent the contribution of multiply scattered light in the dust aggregate. In the absence of multiple scattering, the second terms in the parenthesis become zero; therefore, $(\bar{d}_{1,n}^{(1)}, \bar{d}_{1,n}^{(2)})=(a_n,b_n)$. Thus, the formulation of MFT coincides with the RGD theory.

The harmonic expansions in Equations (\ref{eq:basic1}, \ref{eq:basic2}) are truncated at the order of $\nu_0$. 
We set $\nu_0$ as the closest integer to $x+4x^{1/3}+2$ \citep{B83}.
Using Equations (\ref{eq:basic1}, \ref{eq:basic2}), the mean field $(\bar{d}_{1,n}^{(1)}, \bar{d}_{1,n}^{(2)})$ can be obtained by inversion of a $2\nu_0\times2\nu_0$ complex matrix.
Therefore, the matrix size depends only on the monomer size parameter, and then the computational cost of matrix inversion is irrelevant to the number of monomers. Therefore, optical properties of dust aggregates with a large number of monomers can be readily computed.

Once $(\bar{d}_{1,n}^{(1)}, \bar{d}_{1,n}^{(2)})$ are obtained, one can readily calculate optical properties of fractal dust aggregates. In MFT, the extinction cross section of the dust aggregate is simply the sum of the extinction cross section of the individual monomer; hence, we have
\begin{equation}
C_{\mathrm{ext}}^{(\mathrm{MFT})}=\frac{2\pi N}{k^2}\sum_{n=1}^{\infty}(2n+1)\mathrm{Re}(\bar{d}_{1,n}^{(1)}+\bar{d}_{1,n}^{(2)}) \label{eq:Cextmft},
\end{equation}
where we have used Equations (\ref{eq:Cext} and \ref{eq:rgdmft}) \citep{B97}. Similarly, the scattering cross section is obtained by using Equations (\ref{eq:scargd}, \ref{eq:RGDtheory}, \ref{eq:rgdmft}). The absorption cross section is then obtained by $C_{\mathrm{abs}}^{(\mathrm{MFT})}=C_{\mathrm{ext}}^{(\mathrm{MFT})}-C_{\mathrm{sca}}^{(\mathrm{MFT})}$.

\subsection{Static structure factor} \label{sec:gu}
Computations of the RGD theory and MFT require calculations of the static structure factor. 
The role of the static structure factor is to specify the aggregates structure.
In this paper, we assume that the monomer distribution in the aggregate is statistically isotropic \citep{T16}.
Because the static structure factor is the Fourier transform of the two-points correlation function $g(u)$, it can be written as
\begin{eqnarray}
\mathcal{S}(q)=\int_0^{\infty} g(u)4\pi u^2\frac{\sin(qu)}{qu}du. \label{eq:sq}
\end{eqnarray}
The meaning of $g(u)$ is a probability to find a pair of monomers separated by a distance $u$.
In the case of fractal dust aggregates, the explicit expression of $g(u)$ is available \citep[e.g.,][]{T16}:
\begin{equation}
g(u)=\frac{1}{4\pi R_g^3}\left(\frac{u}{R_g}\right)^{d_f-3}f_c\left(\frac{u}{R_g}\right)+\frac{1}{N}\delta(u), \label{eq:gumodel}
\end{equation}
where $f_c$ represents the cut-off of the two-points correlation function, $d_f$ is the fractal dimension, and $\delta$ is the Dirac delta function.
Note that the radius of gyration and the monomer radius are related via $N=k_0(R_g/R_0)^{d_f}$, where $k_0$ is the fractal prefactor.
In Equation (\ref{eq:gumodel}), the Dirac delta function is inserted so that self-correlation is taken into account \citep{T16}.
Several models of the cut-off function $f_c$ have been proposed so far, such as the exponential cut-off \citep{B86}, Gaussian cut-off \citep[][]{S01,T16}, and fractal dimension cut-off \citep{B95}, and those have the forms of
\begin{eqnarray}
&&f_c\left(\frac{u}{R_g}\right)=\nonumber\\
&&
\left\{ \begin{array}{lll}
    \frac{2^{1-d_f}d_f^{d_f/2}}{\Gamma(d_f/2)}\exp{\left[-\frac{d_f}{4}(u/R_g)^2\right]}, & (\mathrm{GAUSS}) \nonumber\\
   \frac{\{d_f(d_f+1)/2\}^{d_f/2}}{\Gamma(d_f)}\exp{\left[-(\frac{d_f(d_f+1)}{2})^{1/2}(u/R_g)\right]}, & (\mathrm{EXPTL}) \nonumber\\
    \frac{d_f}{2}\exp[{-\frac{1}{2}(u/R_g)^{d_f}}]. & (\mathrm{FLDIM}) \nonumber 
  \end{array} \right.\\ \label{eq:comodel}
\end{eqnarray}
\citet{B95} found that the FLDIM cut-off model seems to be appropriate choice for dust aggregates with $d_f\le2$.
What is the appropriate choice of the cut-off function for dust aggregates with $d_f>2$? 
For $d_f=3$, both the GAUSS and FLDIM models can produce the extinction cross section to an accuracy better than 20\% (see Appendix \ref{sec:app2} for more detail). However, the EXPTL model is not able to reproduce the extinction opacity of BPCA. 

The FLDIM model is slightly more accurate than the GAUSS model. On the other hand, an advantage of the GAUSS model over the FLDIM model is that the computation is fast because integration of Equation (\ref{eq:sq}) can be done analytically \citep{T16}.
In this paper, we use the FLDIM model in computations of the RGD theory and MFT.

\section{Opacity of fractal dust aggregates} \label{sec:opacity}
\subsection{Set up} \label{sec:setup}
We consider two types of fractal dust aggregates, BCCA and BPCA. Typically, BCCA have $d_f=1.9$ and $k_0=1.03$, corresponding to the fluffy structure, and BPCA have $d_f=3.0$ and $k_0=0.3$, corresponding to the compact structure \citep[see also Figure 1 of][]{T16}. 
Although BPCA have $d_f=3$, a porosity of the aggregates is typically high, which is around 86\%. For optical properties of more compact dust aggregates, see e.g., \citet{shen08, shen09, M16}.
We calculate optical properties of BCCA and BPCA using TMM.
Calculations of TMM are performed using the Fortran 77 code SCSMFO1B.FOR\footnote{The code is available on \url{ftp://ftp.eng.auburn.edu/pub/dmckwski/scatcodes/}. A newer version of this code is available at \url{http://www.eng.auburn.edu/~dmckwski/scatcodes/}.} \citep{M96} armed with the Quasi Monte Carlo method \citep{O08}. The Quasi Monte Carlo method is a method that enables fast computation of orientation averaging of dust aggregates.
In our TMM calculations, optical properties of each aggregate is averaged over 30 different orientations. In addition, we produce 10 (4) different BCCA (BPCA) with the same number of monomers, and then their optical properties are averaged.
We use the optical constants of astronomical silicate \citep{D84, L93}, and the material density is assumed to be $3.3$ g cm$^{-3}$.

Denote $N_O$ by the number of orders retained in the harmonic expansions of each monomer. The required value of $N_O$ for convergence depends on the refractive index and the proximity of neighboring monomers \citep{Mackowski:1994aa}. 
Because astronomical silicate has large values of the refractive index at FIR and mm wavelengths \citep[see e.g., Figure 3 of][]{T16}, large $N_O$ value is required for convergence even if the monomer radius is much smaller than the wavelength \citep[see e.g.,][]{Mackowski:1994aa, Mackowski:1995aa, Mackowski:2006aa}.
In order to attain convergence at FIR and mm wavelengths for astronomical silicate, we set $N_O=10$ for $\lambda>10\ \mu$m. For the case of a two-sphere cluster, $N_O=10$ gives rise to the relative error of the extinction cross section less than 0.35\%. For $\lambda<10\ \mu$m, we truncate the harmonic expansion of each monomer particle until the error torrence of the extinction efficiency of $10^{-4}$ is attained.

\subsection{Validity of MFT and the RGD theory} \label{sec:wavel}
\begin{figure*}[t]
\begin{center}
\includegraphics[height=13.5cm,keepaspectratio]{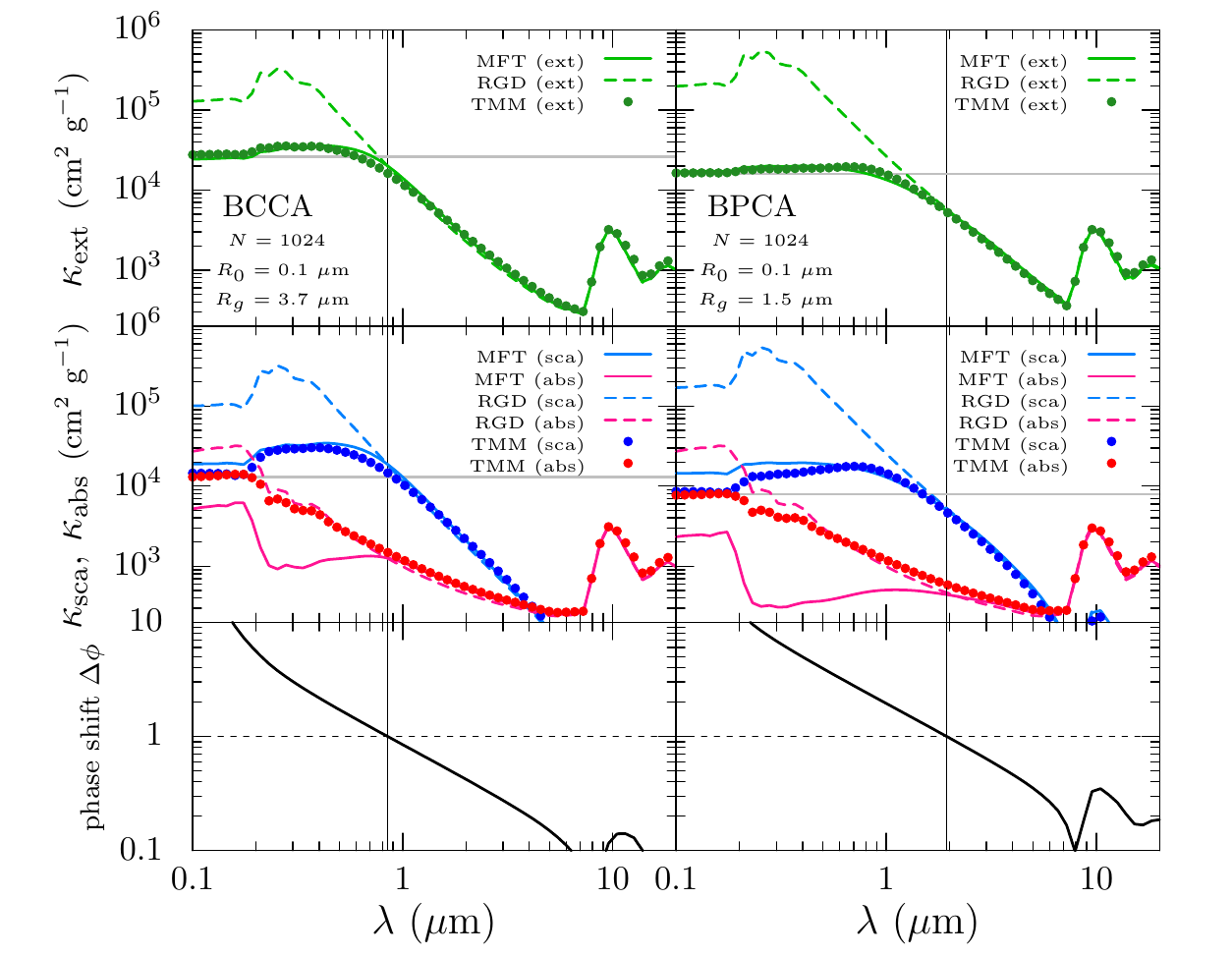}
\caption{ Extinction opacity (top), scattering and absorption opacities (middle), and the phase shift $\Delta\phi$ (bottom). 
Left and right panels show opacities of BCCA and BPCA of 1024 monomers with $0.1\ \mu$m radii, respectively. 
(Top and middle panels): Dots, solid, and dashed lines represent the results obtained using TMM, MFT, and the RGD theory. Blue and red colors indicate scattering and absorption opacities, respectively. Gray horizontal lines is the opacity value expected from the geometrical optics approximation. Black vertical lines running from bottom to top panels are the wavelength where the phase shift becomes unity.}
\label{fig:f1}
\end{center}
\end{figure*}

We calculate the opacities of BCCA and BPCA using TMM, the RGD theory, and MFT. The monomer radius and the number of monomers are set as $R_0=0.1\ \mu$m and $N=1024$, respectively. Hence, BCCA and BPCA have the radii of gyration of $R_g=3.73\ \mu$m and $R_g=1.51\ \mu$m, respectively \citep[see Table 1 in][]{T16}. 
In Figure \ref{fig:f1}, we show the wavelength dependence of opacities of the BCCA and BPCA models.

First of all, the extinction opacities of the BCCA and BPCA models are discussed.
At $\Delta\phi<1$ (longer wavelength domain), the RGD theory can reproduce the extinction opacities of both BCCA and BPCA models.
On the other hand, at $\Delta\phi>1$, the opacities obtained by the RGD theory are significantly overestimated. 
This is because at $\Delta\phi>1$, multiple scattering cannot be negligible, and hence the single scattering assumption of the RGD theory breaks down. 
As a result, it is found that the validity of the RGD theory can be well predicted by Equation (\ref{eq:rgdcond2})\footnote{\citet{T16} used double inequality in their Equations (5, 6, and 7) as the condition of the RGD theory. However, from Figure \ref{fig:f1}, we conclude that the double inequalities are too conservative, and hence, the double inequalities in their Equations (5, 6, 7) should be replaced by single inequalities to represent more realistic conditions of the RGD theory.}.
In order to obtain correct opacities at $\Delta\phi>1$, we use MFT, where multiple scattering can be taken into account.
Surprisingly, MFT can reproduce the extinction opacities of both the BCCA and BPCA models.
The extinction opacities obtained by MFT agree with the TMM results to an accuracy of $\leq 25\%$ and $\leq15\%$ for both BCCA and BPCA models, respectively. 

Secondly, we discuss the scattering and absorption opacities of BCCA and BPCA models. As we have discussed so far, MFT can account for the extinction opacity well. However, as shown in Figure \ref{fig:f1}, the scattering and absorption opacities obtained by MFT deviate from the TMM results when $\Delta\phi>1$. This reason is discussed in Section \ref{sec:interprete} and Appendix \ref{sec:BP}.

Finally, we mention convergence of the TMM results to the geometrical optics limit at the short wavelength domain.
Horizontal solid lines in Figure \ref{fig:f1} indicate the opacity in geometrical optics limit, where the geometrical cross sections are obtained by using the empirical formulae derived by \citet{M06}. Note that, in the case of extinction, the opacity is a twice of the geometrical cross section divided by the mass of the aggregate. It can be seen from Figure \ref{fig:f1} that when $\Delta\phi\gtrsim10$, the extinction opacities obtained by TMM agree with those obtained from the geometrical optics limit.

\subsection{Interpretations of the MFT results} \label{sec:interprete}
Figure \ref{fig:f1} shows that MFT gives reliable results for the extinction opacity at arbitrary $\Delta\phi$, but it fails to reproduce the scattering and absorption opacities at $\Delta\phi>1$. The incorrect behavior of MFT is due to the fact that the single scattering albedo at multiple scattering domain obtained by MFT is exactly the same as that of the RGD theory, which is a single scattering theory (see Appendix \ref{sec:BP} for more detail). 

\begin{figure}
\begin{center}
\includegraphics[height=6.0cm,keepaspectratio]{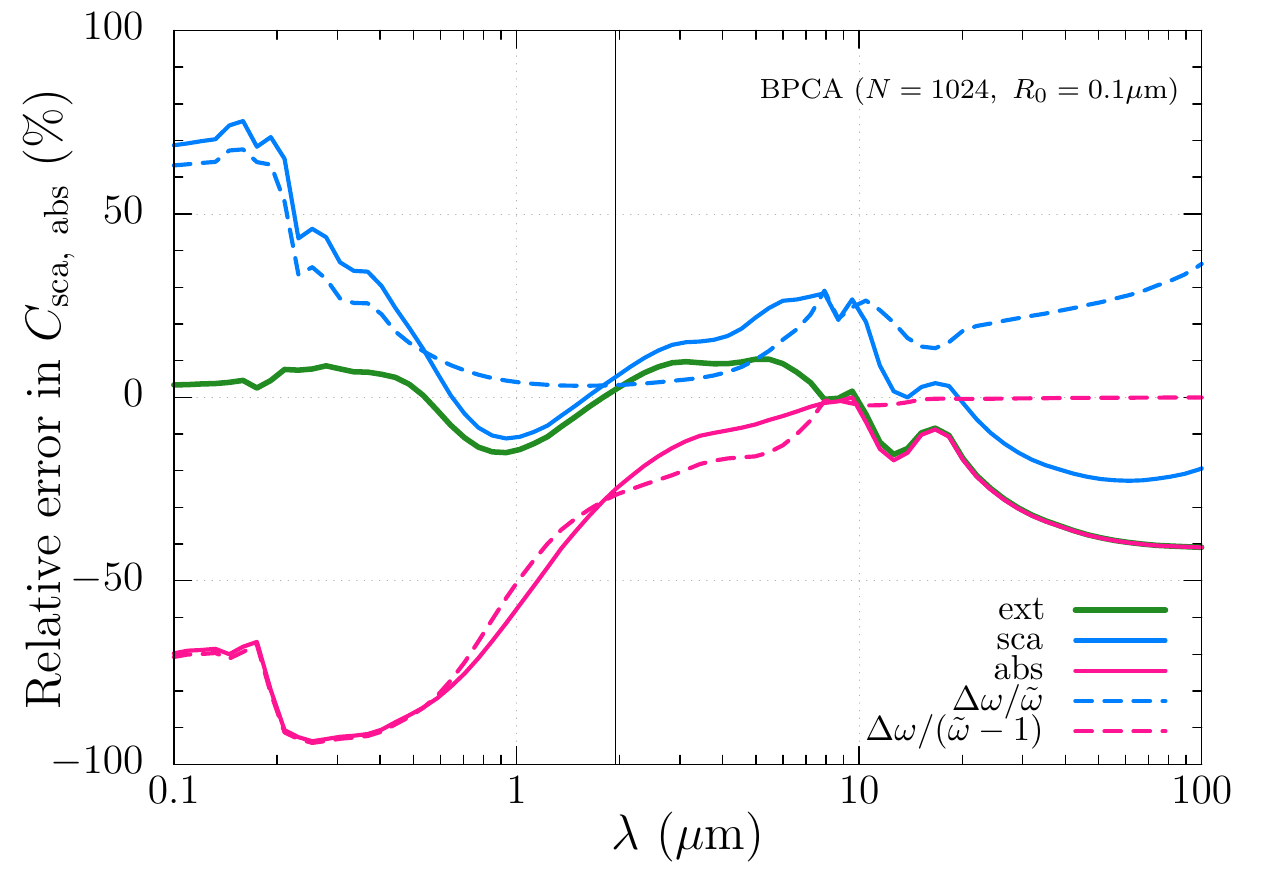}
\caption{Relative errors of the MFT in extinction, scattering, and absorption cross sections of BPCA  (solid lines). Green, blue, and red lines correspond to the extinction, scattering and absorption cross sections, respectively. Dashed lines represent the error of the scattering and absorption cross sections estimated from the error of the single scattering albedo. The vertical line denotes the wavelength where the phase shift of the aggregate becomes unity.}
\label{fig:error}
\end{center}
\end{figure}

In order to demonstrate that the error of single scattering albedo correlates with the incorrect behavior of the scattering and absorption opacities, we show the relative errors of MFT in Figure \ref{fig:error}. 
Let $\Delta\omega$ denote the residual in the single scattering albedo between the RGD theory and the TMM results; hence, we can write $\Delta\omega\equiv\omega^{(\mathrm{RGD})}-\omega^{(\mathrm{TMM})}$, where $\omega^{(\mathrm{RGD})}$ and $\omega^{(\mathrm{TMM})}$ represent the single scattering albedo calculated by the RGD theory and TMM.
Since the RGD theory tends to overestimate the albedo, $\Delta\omega>0$. Suppose the extinction cross section obtained by MFT is correct, then the relative errors in the scattering and absorption opacities of MFT can be estimated to be $\Delta_{\mathrm{sca}}\simeq\Delta\omega/\omega^{(\mathrm{TMM})}$ and $\Delta_{\mathrm{abs}}\simeq\Delta\omega/(\omega^{(\mathrm{TMM})}-1)$, respectively. 

In Figure \ref{fig:error}, it is clear that the incorrect behavior of the single scattering albedo of the RGD theory well correlates with the relative error of the absorption and scattering opacities obtained by MFT. The absorption cross section (or opacity) will be underestimated by at most 94\% for BPCA and 87\% for BCCA with $N=1024$ and $R_0=0.1\ \mu$m. 
As a result, it is shown that the over/under-estimation of the scattering and absorption opacities is due to the fact that the albedo of MFT is the same as that of the RGD theory. This property of MFT is due to the mean field assumption; therefore, the over/under-estimation of the scattering and absorption cross sections is inevitable when MFT is used.

Figure \ref{fig:error} also shows that the error in the extinction cross section becomes large at long wavelength domain. In addition, it is also found that this error does not correlate with that of the single scattering albedo. This is due to the interaction of neighboring monomers which is important when the refractive index is large at the Rayleigh domain (see also Sections \ref{sec:syserror} and \ref{sec:excess})

\subsection{Modified Mean Field theory} \label{sec:mmf}
As shown in Sections \ref{sec:wavel} and \ref{sec:interprete}, MFT fails to reproduce the scattering and absorption opacities when multiple scattering is important. In order to avoid these problems, we propose a prescription which enables to calculate the opacities at multiple scattering domain almost correctly. Hereafter, we call our approach as a modified mean field theory (MMF).

The opacity calculation in MMF is performed by as follows.
First of all, we calculate the extinction cross section using MFT (Equation \ref{eq:Cextmft}), that is, $C_{\mathrm{ext}}^{\mathrm{(MMF)}}=C_{\mathrm{ext}}^{\mathrm{(MFT)}}$, since MFT successfully reproduces the extinction cross section as shown in Section \ref{sec:wavel}.
The second step is to calculate the absorption cross section.
As we have mentioned, the cross section obtained by the TMM results approach to the geometrical optics limit at the short wavelength domain. In the geometrical optics limit, the absorption and scattering cross sections can be given by the geometrical cross section of dust aggregates unless the reflection comes into play \citep[e.g.,][]{B83}. 
Based on above considerations, we introduce the following empirical expression
\begin{eqnarray}
C_{\mathrm{abs}}^{\mathrm{(MMF)}}&=&\mathrm{max}\left(C_{\mathrm{abs}}^{(\mathrm{MFT})}, G(1-e^{-\tau})\right),\label{eq:mmf}\\
\tau&=&C_{\mathrm{abs}}^{(\mathrm{RGD})}/G, \label{eq:tau}
\end{eqnarray}
where $G$ is the geometrical cross section of the aggregates.  
By using Equation (\ref{eq:mmf}), the absorption cross section obtained by MFT is connected to the geometrical cross section at the short wavelength domain. When the absorption cross section obtained by the RGD theory exceeds the geometrical cross section ($\tau\gg1$), the absorption cross section becomes equal to the geometrical cross section $G$. In the long wavelength domain, $\tau\ll1$, and then $C(1-e^{-\tau})=C_{\mathrm{abs}}^{(\mathrm{RGD})}$.
Note that in Equation (\ref{eq:mmf}), the maximum value is taken so that MFT is chosen at the single scattering region rather than the RGD theory. 
The geometrical cross section $G$ is estimated using the empirical formulae of \citet{M06}. 
Finally, we calculate the scattering cross section using $C_{\mathrm{sca}}^{\mathrm{(MMF)}}=C_{\mathrm{ext}}^{\mathrm{(MMF)}}-C_{\mathrm{abs}}^{\mathrm{(MMF)}}$.

\begin{figure*}
\begin{center}
\includegraphics[height=6.0cm,keepaspectratio]{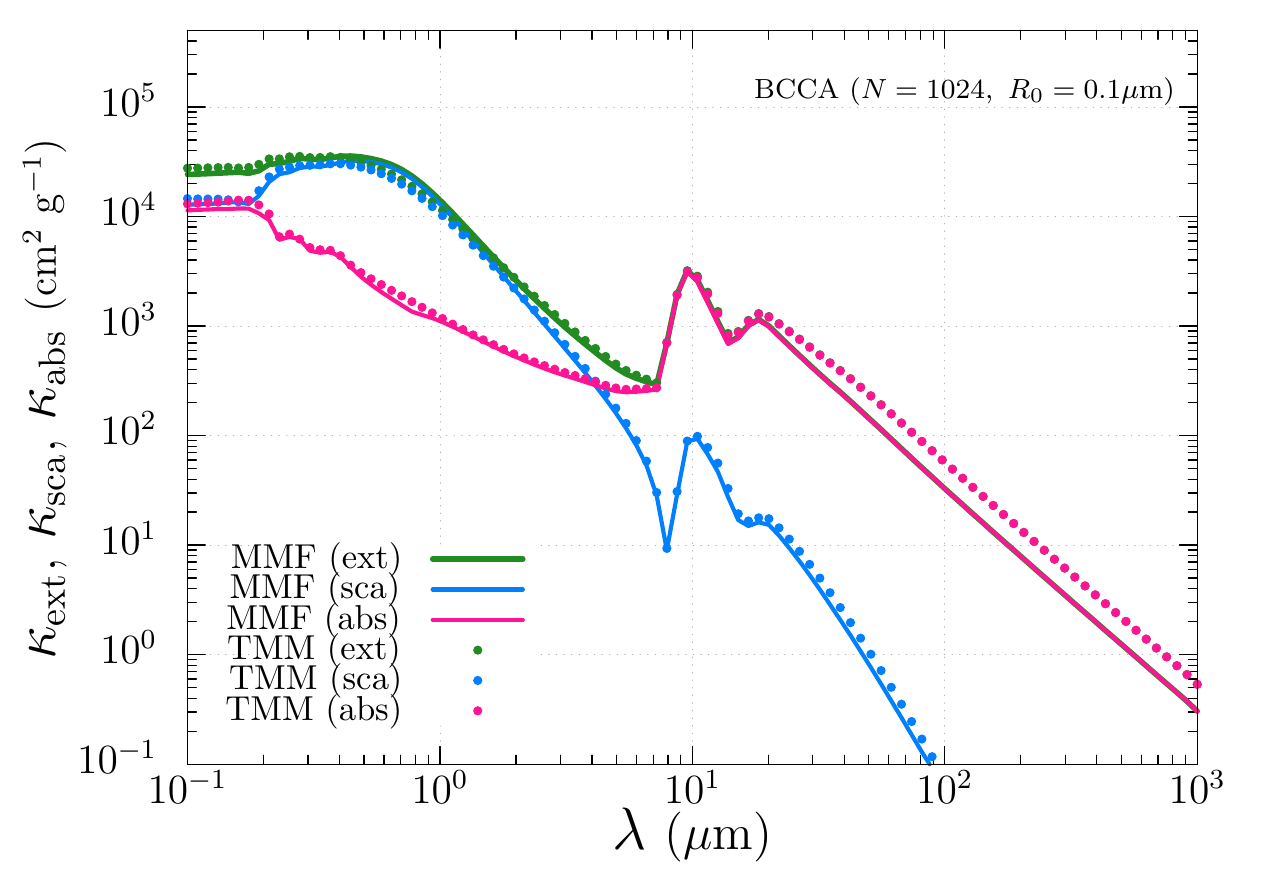}
\includegraphics[height=6.0cm,keepaspectratio]{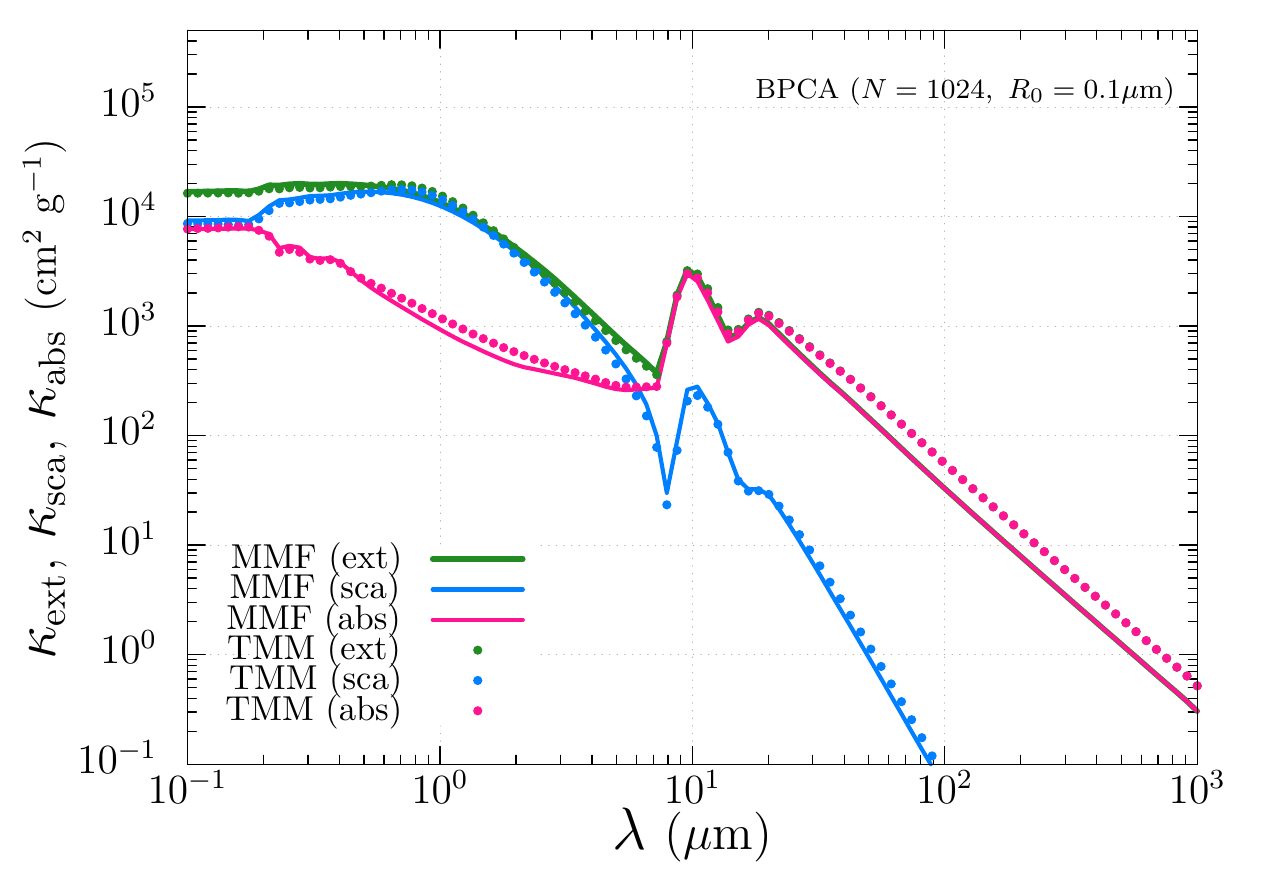}
\caption{Opacities obtained by MMF. Left and right panels correspond to the BCCA ($N=1024$, $R_0=0.1\ \mu$m) and BPCA ($N=1024$, $R_0=0.1\ \mu$m) models, respectively. Green, blue, and red colors indicate extinction, scattering, and absorption opacity, respectively. The TMM and the MMF results are shown in dots and solid lines, respectively.}
\label{fig:mmfsum}
\end{center}
\end{figure*}

\begin{figure}
\begin{center}
\includegraphics[height=6.0cm,keepaspectratio]{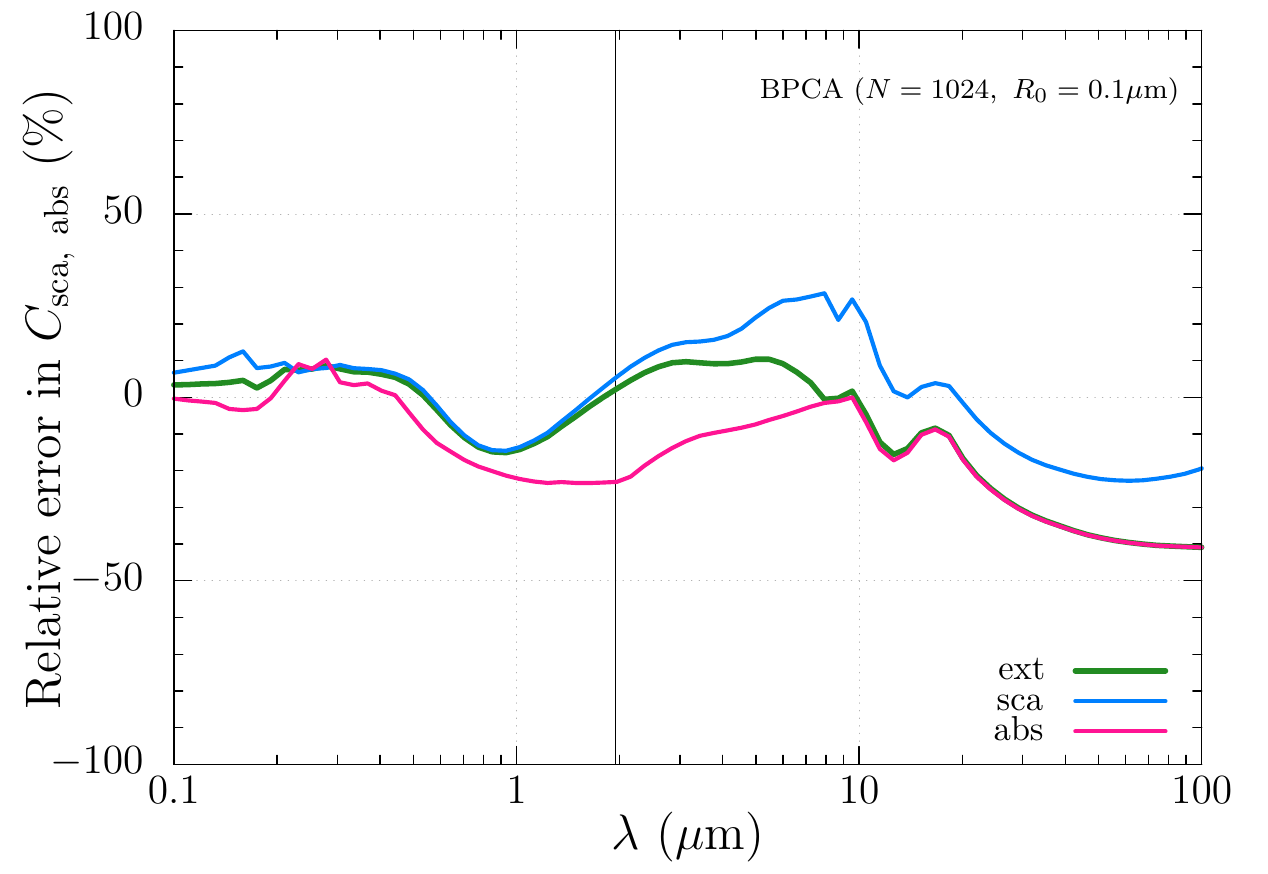}
\caption{Same as Figure \ref{fig:error}, but for MMF.}
\label{fig:error2}
\end{center}
\end{figure}

In Figure \ref{fig:mmfsum}, the opacities calculated by MMF for the BCCA and BPCA models are shown. In addition, Figure \ref{fig:error2} shows the relative error in extinction, absorption, and scattering cross sections of MMF and TMM. The absorption opacities of MMF agree with the TMM results to an accuracy of $\leq 20\%$ and $\leq 25\%$ at $\lambda\le 20\ \mu$m for BCCA and BPCA, respectively. Since MFT gives rise to the errors in the absorption opacity as large as $87\%$ and $94\%$ for the BCCA and BPCA models, respectively, the MMF can improve the accuracy of the absorption opacity. In addition, accuracy of the scattering opacity is also improved in the MMF. 
One may notice that the absorption opacity is systematically underestimated at the long wavelength domain.
In this region, the absorption opacities are underestimated by 41\% and 43\% for the BPCA and BCCA models, respectively (see Sections \ref{sec:syserror} and \ref{sec:excess} for more detail). 

\begin{figure*}
\begin{center}
\includegraphics[height=13cm,keepaspectratio]{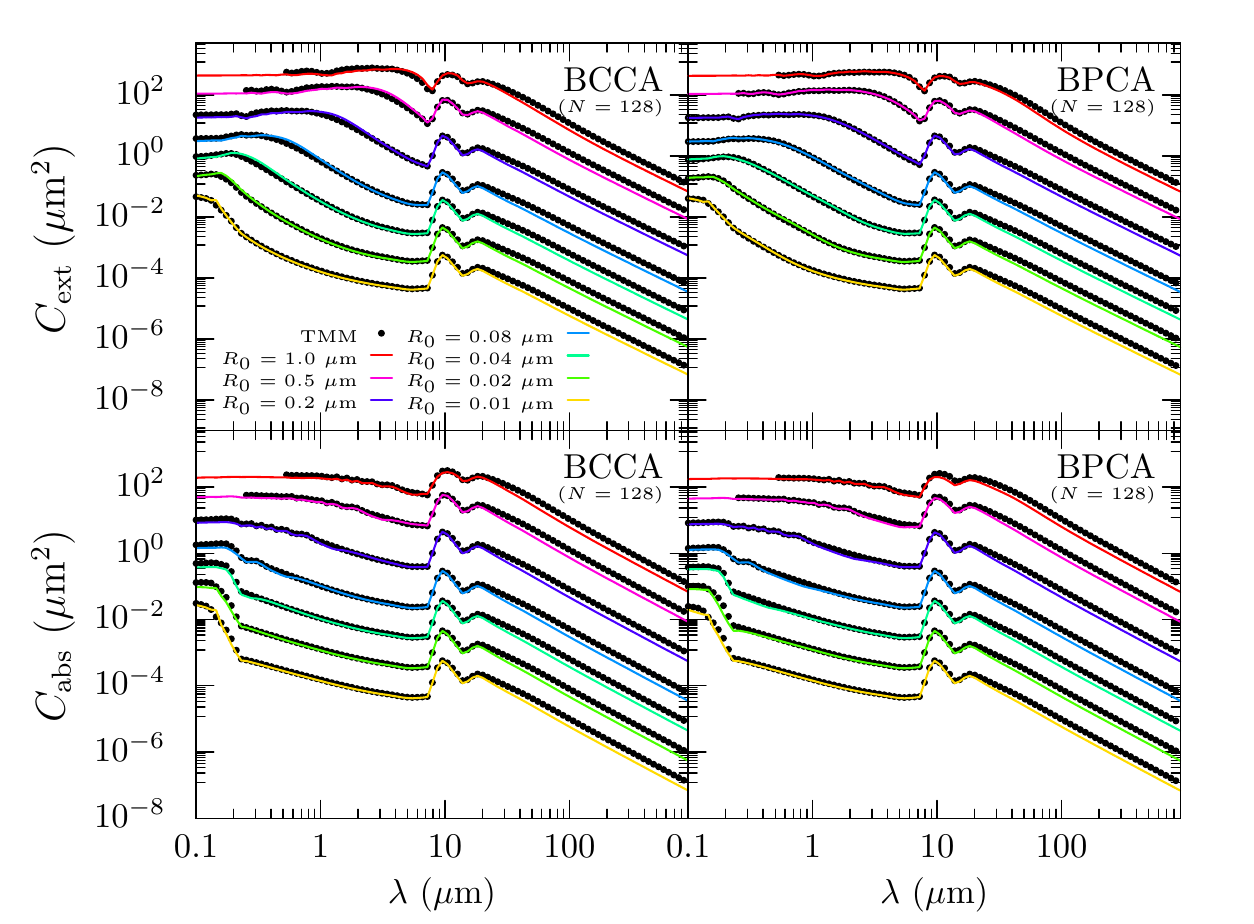}
\caption{Extinction and absorption cross sections of BCCA ($N=128$) and BPCA ($N=128$). The TMM results are denoted by the black dots, and solid lines with different colors indicate the opacity obtained by the MMF with different monomer radii.}
\label{fig:doya}
\end{center}
\end{figure*}

Next, in order to check the validity of MMF at various parameters, we vary the monomer radius. 
Figure \ref{fig:doya} shows the extinction and absorption cross sections of BCCA and BPCA obtained by TMM and MMF, where the number of monomers is set as $N=128$ and the monomer radii are varied from $R_0=0.01\ \mu$m to $R_0=1.0\ \mu$m.
To reduce computational costs, we use $N=128$ instead of $N=1024$. 
Although at the long wavelength domain, MMF systematically underestimates the absorption (or extinction) opacity, MMF can reproduce overall wavelength dependence of the opacities. Even if the monomer radius is varied, MMF still works well. As a result, although the empirical prescription adopted in MMF is simple, accuracy of the absorption cross sections of MFT is significantly improved. 

\subsection{Error at the long wavelength domain} \label{sec:syserror}
In Figures \ref{fig:mmfsum} and \ref{fig:doya}, it was shown that MMF underestimates the absorption opacity at the long wavelength domain. At this wavelength domain, the RGD theory, MFT, and MMF show almost the same results, since the single scattering is the dominant ($\Delta\phi<1$). Hence, this systematic underestimation is a common property in light scattering models presented in Section \ref{sec:models}. 
This error is due to the proximity effect of neighboring monomers, and this is related to the violation of Equation (\ref{eq:rgdcond1}).
Astronomical silicate has large refractive index at long wavelength domain (see Figure 3 in \citet{T16}). 
According to the results shown in Figure \ref{fig:mmfsum} for the BPCA model, relative error of MMF exceeds 20\% at $|m-1|>2$. 
Hence, satisfying Equation (\ref{eq:rgdcond1}) guarantees the accuracy better than 20\% in the RGD theory, MFT, and MMF. The physical origin of this effect, or the proximity effect, is further discussed in Section \ref{sec:excess}.

\section{Detailed comparisons to other approximate methods} \label{sec:approx}
We compare the opacities and the asymmetry parameter obtained by TMM to those obtained by EMT and DHS.
In addition, we also compare the opacities of porous dust aggregates and a compact sphere with the volume equivalent radius, $r_V=R_0N^{1/3}\simeq 1.0\ \mu$m.
Note that all acronyms ``EMT" used in this section denote EMT (MG-Mie).

\begin{figure*}
\begin{center}
\includegraphics[height=6.0cm,keepaspectratio]{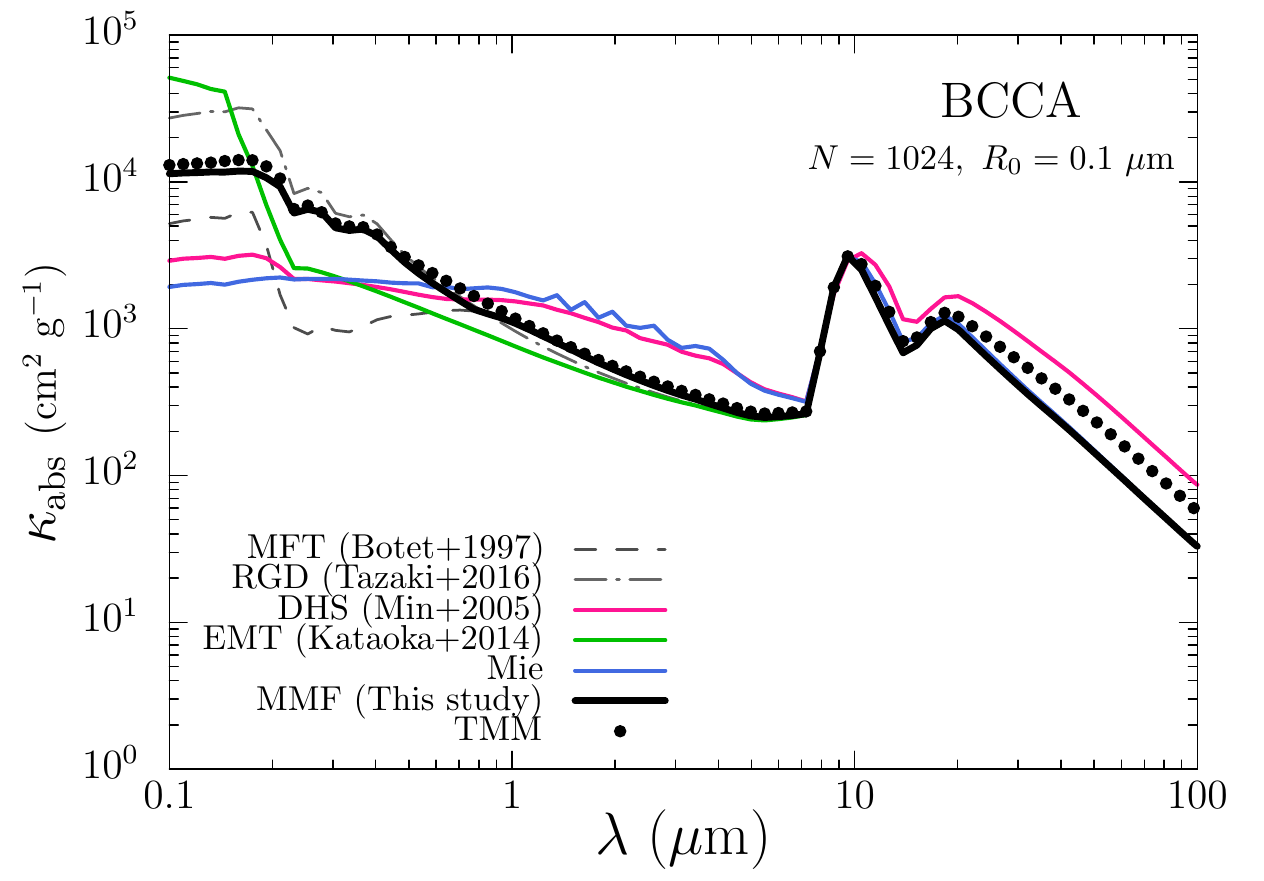}
\includegraphics[height=6.0cm,keepaspectratio]{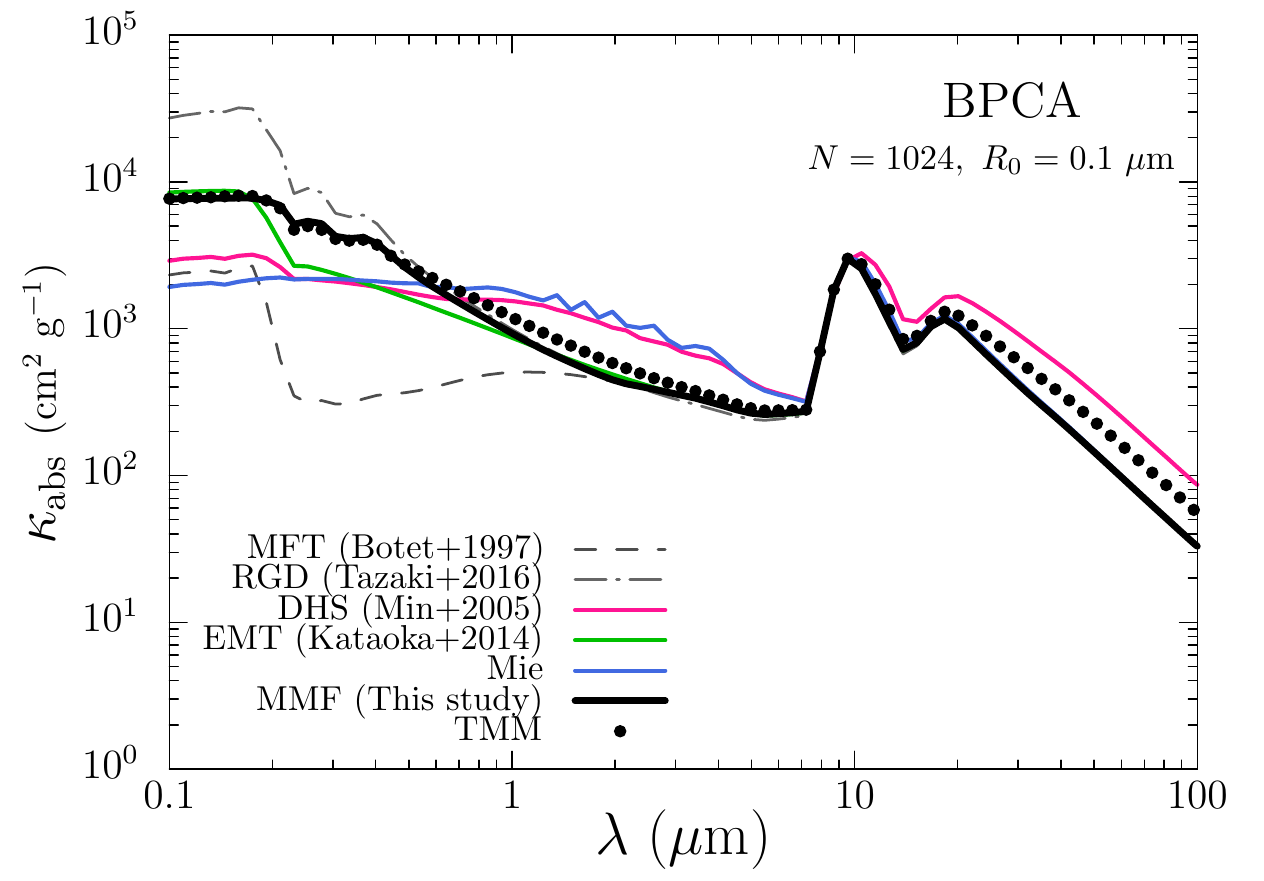}
\includegraphics[height=6.0cm,keepaspectratio]{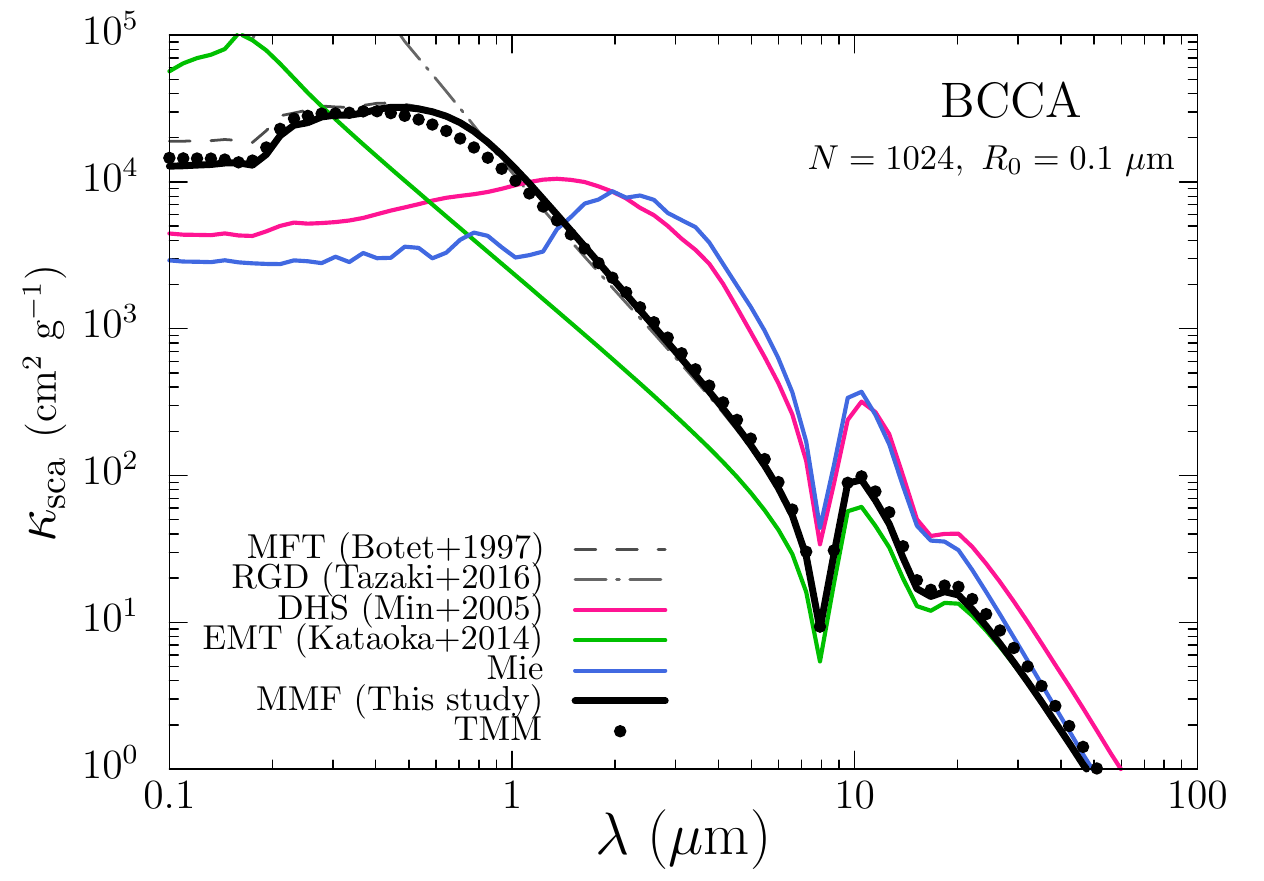}
\includegraphics[height=6.0cm,keepaspectratio]{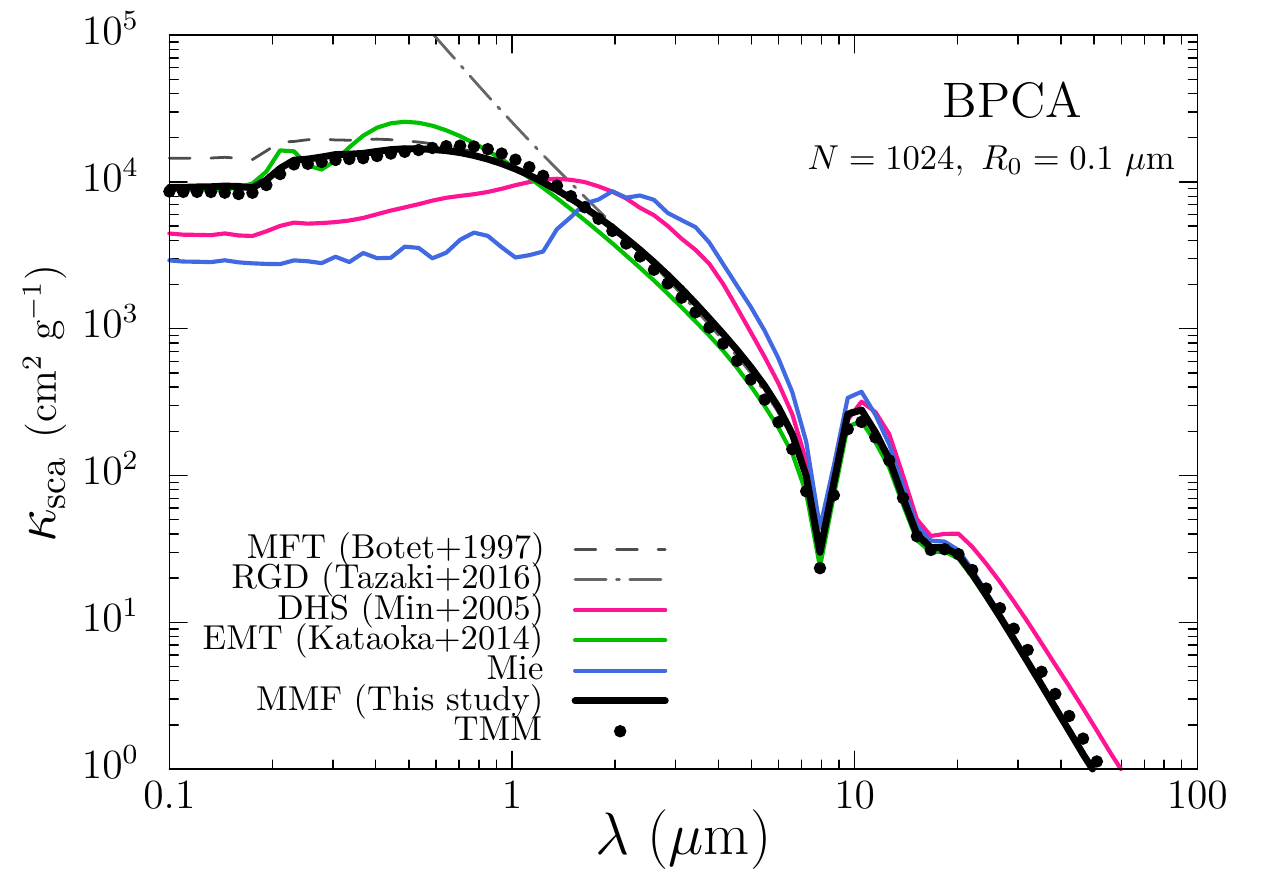}
\includegraphics[height=6.0cm,keepaspectratio]{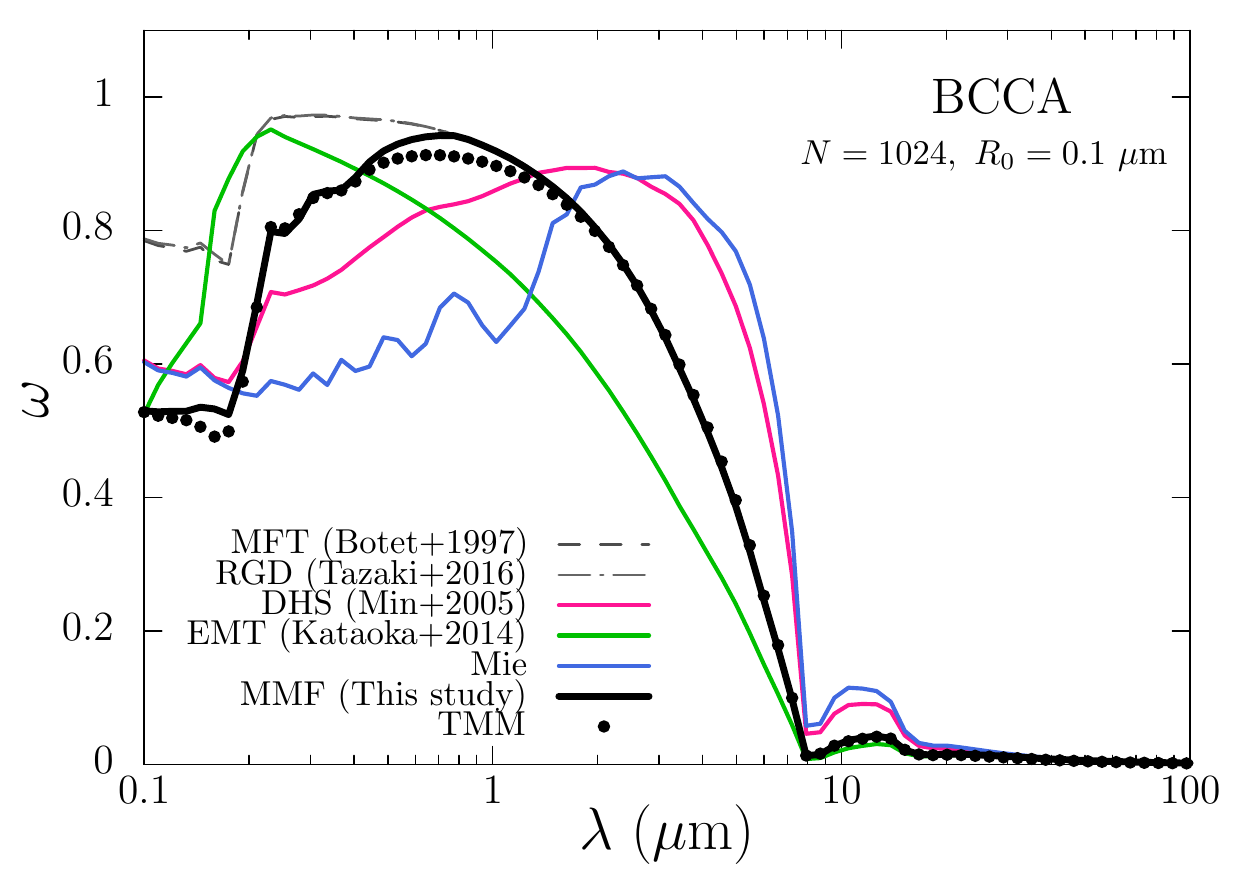}
\includegraphics[height=6.0cm,keepaspectratio]{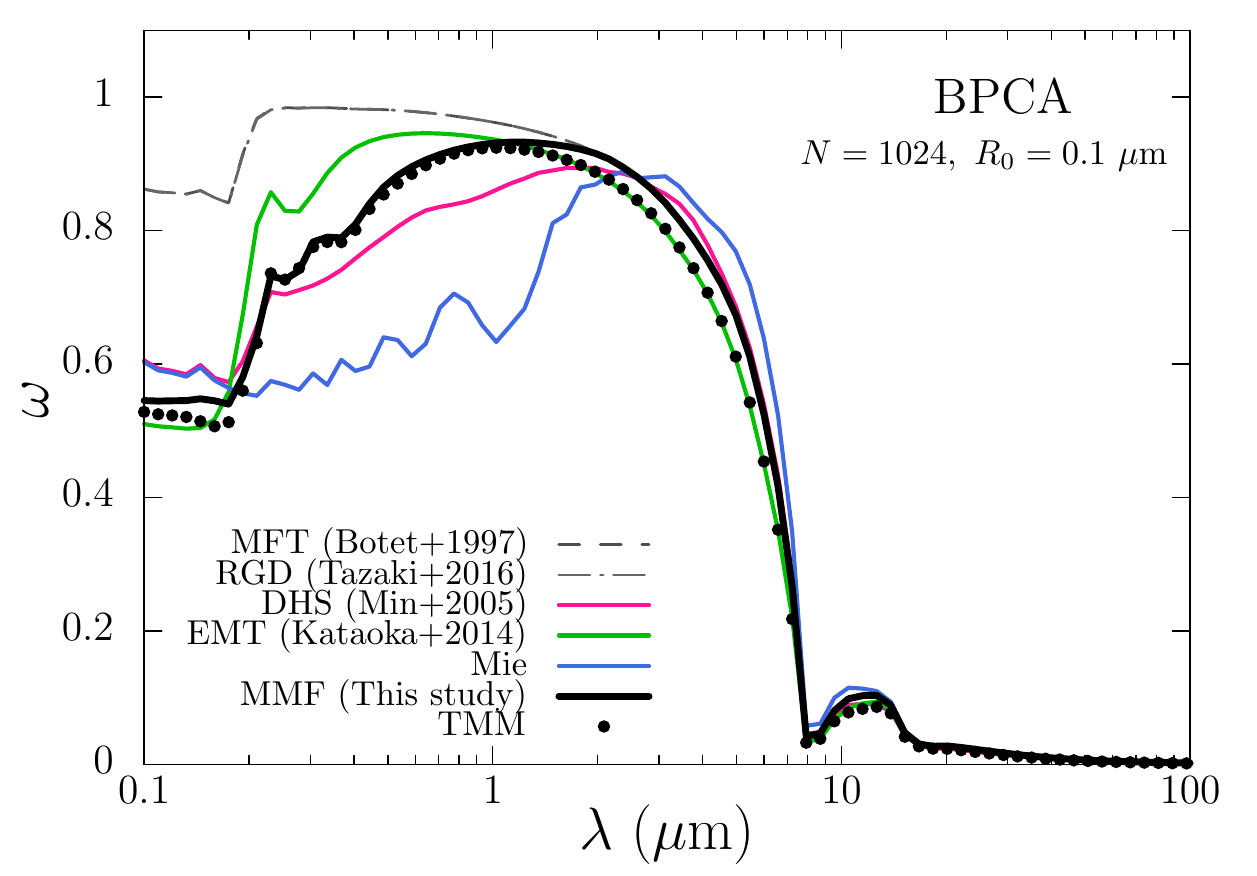}
\caption{Absorption and scattering opacities (top and middle) and single scattering albedo (bottom). Left and right columns correspond to the BCCA and BPCA models with $N=1024$ and $R_0=0.1\ \mu$m. Dots represent the results obtained by TMM, and lines represent those obtained by approximate methods, MFT (gray dashed line), RGD (gray dot-dashed line), DHS (magenta solid line), EMT (MG-Mie) (green solid line), Mie (blue solid line), and MMF (black solid line).}
\label{fig:kappa1}
\end{center}
\end{figure*}

We calculate the absorption and scattering opacities of BCCA and BPCA with $N=1024$ and $R_0=0.1\ \mu$m, and the results are presented in Figure \ref{fig:kappa1}. Detailed comparisons of the results between TMM and the Mie theory, EMT, and DHS are given in Sections \ref{sec:mietmm}, \ref{sec:emttmm}, and \ref{sec:dhstmm}, respectively.
The asymmetry parameter and the effective albedo of the aggregates are discussed in Section \ref{sec:compare2}.

\subsection{Opacities} \label{sec:compare1}

\subsubsection{Comparison between the Mie theory and TMM} \label{sec:mietmm}
As a starting point of our discussion, it is helpful to see how porous dust aggregates (TMM) differ from a homogeneous sphere with the volume equivalent radius (the Mie theory) in the opacities.
In Figure \ref{fig:kappa1}, we plot the results obtained using the Mie theory for a sphere with radius $r_V=R_0N^{1/3}$, where the optical constant is the same as that of the monomer, that is, astronomical silicate. 

First of all, we discuss the absorption opacities (top panels in Figure \ref{fig:kappa1}). 
In Figure \ref{fig:kappa1} (top panels), it is clear that the absorption opacities obtained by the Mie theory are significantly different from those obtained by TMM for both BCCA and BPCA models. 
When $\lambda\gtrsim 2\pi r_V\simeq 6\ \mu$m, the Mie theory underestimates the absorption opacities. 
This is because the Mie theory cannot correctly treat the proximity effect of the monomers (see Section \ref{sec:excess} for more detail). 
This indicates that even if the wavelength is much longer than the radius of the dust aggregate, the homogeneous sphere with the volume equivalent radius is not a good representative of the porous dust aggregate in terms of opacities. 
This is mainly due to the fact that when $|m-1|\gg1$, the absorption opacity in the Rayleigh domain is very sensitive to particle shape.
Note that in the Rayleigh domain, the Mie theory with a volume equivalent sphere gives the absorption cross section of $NC_{\mathrm{abs}}^0$, which is the same as that of MMF\footnote{In the long wavelength domain, single scattering is dominant ($\Delta\phi<1$), and hence, the absorption cross section of MMF and MFT becomes the same as that of the RGD theory.}.
At $\lambda\lesssim 2\pi r_V\simeq 6\ \mu$m, the opacities obtained by the Mie theory show shallower wavelength dependence than those obtained by TMM.
The absorption opacities of the BPCA and BCCA models are similar to that of a single monomer as long as $\Delta\phi<1$ (see, for example, lines of the RGD theory in top panels of Figure \ref{fig:kappa1}), whereas the absorption opacity of the Mie theory reflects the property of the particle larger than the monomer radius.
As a result, the Mie theory tends to overestimate the absorption opacities at infrared wavelengths, whereas it underestimates the opacities at visible and UV wavelengths. 

Next, we discuss the scattering opacities of the BCCA and BPCA models (middle panels in Figure \ref{fig:kappa1}). In the case of scattering, the structure of dust aggregates plays essential roles in the scattering process \citep{T16}. The Mie theory deviates from the TMM results at $\lambda \lesssim 2\pi R_g$. Since BCCA and BPCA models show $R_g=3.7\ \mu$m and $R_g=1.5\ \mu$m, these wavelengths are corresponding to $\lambda\simeq\ 23\ \mu$m and $\lambda\simeq\ 9.4\ \mu$m for the BCCA and BPCA models, respectively. As a result, the Mie theory cannot predict the scattering opacity outside the Rayleigh domain of the dust aggregates ($\lambda\lesssim 2\pi R_g$). 

Since the Mie theory fails to reproduce both absorption and scattering opacities of dust aggregates, the single scattering albedo also shows incorrect behavior at the short wavelength domain (bottom panels in Figure \ref{fig:kappa1}). Note that as the wavelength decreases, the single scattering albedo obtained by both the Mie theory and TMM approaches to a value of 0.5. This value corresponds to the single scattering albedo in the geometrical optics limit without reflection.

Therefore, the Mie theory with the volume equivalent radius cannot capture the opacity at both inside and outside the Rayleigh domain of dust aggregates.

\subsubsection{Comparison between EMT (MG-Mie) and TMM} \label{sec:emttmm}
In the EMT method, a dust aggregate is approximated as a single homogeneous sphere with the effective dielectric function, and then the Mie theory is used to obtain optical properties \citep[e.g.,][]{Mukai:1992aa}. The effective dielectric function is commonly derived using either the Maxwell--Garnett mixing rule or the Bruggeman mixing rule, although various mixing rules have so far been proposed \citep{Ossenkopf91, Stognienko:1995aa, Henning96, Min:2008aa}. 
\citet{vosh07} compared different mixing rules and found that the Bruggeman mixing rule generally provides reliable results.
When the volume fraction of the inclusion is much smaller than that of matrix, the Maxwell-Garnett mixing rule can also give reliable results. Because we consider porous dust aggregates (porosity $\gtrsim 85\%$), we adopt the Maxwell--Garnett mixing rule \citep{MG09}.
The filling factor of the dust aggregate is defined by using its characteristic radius \citep{Mukai:1992aa}.

In Figure \ref{fig:kappa1}, we compare the wavelength dependence of EMT (MG-Mie) and TMM.

First of all, we discuss the absorption opacities (top panels in Figure \ref{fig:kappa1}). 
Similar to the case of the homogeneous sphere, EMT also fails to reproduce absorption opacity at the Rayleigh domain ($kR_c<1$). 
This is because EMT (MG-Mie) cannot reproduce the proximity effect of neighboring monomers (see Section \ref{sec:excess} for more detail). Note that in the Rayleigh domain, the absorption cross section calculated by EMT (MG-Mie) equals to $NC_{\mathrm{abs}}^0$, which is the same as that of MMF.

At $kR_c>1$, EMT seems to reproduce the absorption opacities for both aggregates models except for $\lambda \lesssim 1\ \mu$m. Using EMT (MG-Mie), \citet{K14} showed that the absorption opacities of the dust aggregates are the same as long as $R_cf$ remains constant. Since $R_cf\propto N^{1-2/d_f}$, the absorption opacity of the dust aggregates with $d_f=2$, like BCCA, does not depend on the number of monomers $N$. Thus, the absorption opacity of BCCA obtained by EMT becomes almost the same as that of the single monomer. 
Note that when the aggregate radius is not too large, the absorption opacity of dust aggregates with $d_f=3$ becomes also similar to that of the single monomer.
As a result, for both BCCA and BPCA models, EMT shows the absorption opacity similar to that of the monomer, and hence, EMT seems to reproduce the opacity obtained by TMM. However, once the wavelength becomes close to the monomer radius, e.g., $\lambda\lesssim 1\ \mu$m, the EMT results deviate from the TMM results. This deviation is due to the fact that the Maxwell-Garnet mixing rule assumes that the monomers are in the Rayleigh domain; thus, it cannot be used where the monomer size parameter is large.

Next, we discuss the scattering opacity (middle panels in Figure \ref{fig:kappa1}). 
EMT is able to reproduce the scattering opacity of BPCA at almost all wavelengths. 
The opacity of BPCA is characterized by small angle scattering. Since EMT is able to reproduce small angle scattering almost correctly \citep[see Figure 5(a) of][]{T16}, EMT successfully reproduce the scattering opacity of BPCA. However, EMT fails to reproduce the scattering opacity of BCCA when $\lambda\lesssim 2\pi R_g$. 
In the case of BCCA, scattering opacity is characterized not only small angle scattering but also intermediate angle scattering. EMT significantly underestimates the scattered intensity at intermediate and back scattering angles \citep[see Figure 5(a) of][]{T16}; therefore, EMT fails to reproduce the scattering opacity of the BCCA model when $\lambda\lesssim 2\pi R_g$.

Thirdly, we discuss the single scattering albedo.
Since EMT cannot reproduce the scattering opacity outside the Rayleigh domain for BCCA, the single scattering albedo is also incorrect at this region. For BPCA, the single scattering albedo is almost accurate unless the monomer size parameter is large.

Our results imply that optical properties of porous dust aggregates are governed not only porosity, but also fractal dimension. Even if the porosity of two dust aggregates are similar, their optical properties can be different when their fractal dimensions are not the same.
\citet{vosh07} performed the DDA calculations of porous spheres and concluded that EMT can be applicable to highly fluffy dust aggregates (porosity $> 90\%$) if the mixing rule is properly selected. However, the particle configuration adopted in the DDA calculations in \citet{vosh07} is very similar to the randomly distributed Rayleigh inclusions encased in a given spherical volume. It is known that such a particle configuration gives rise to show similar optical properties to those of EMT \citep[see e.g.,][]{Mishchenko:2016aa}. Therefore, the agreement between DDA and EMT in \citet{vosh07} is presumably due to their particle configuration. Indeed, the phase function of fractal dust aggregates of Rayleigh monomers (porosity $> 90\%$) is significantly different from that obtained by EMT \citep{kozasa93, shen09, T16}.

\subsubsection{Comparison between DHS and TMM} \label{sec:dhstmm}
DHS is an approximate method to obtain scattering and absorption properties of irregularly shaped particles \citep{M03, M05, M16}.
In DHS, optical properties of hollow spheres are averaged over a distribution, while keeping the material volume preserved:
\begin{eqnarray}
\langle A \rangle_{\mathrm{DHS}}&=&\int_0^1 n(f_{\mathrm{v}})A(r_{\mathrm{in}},r_{\mathrm{out}})df_{\mathrm{v}}, \label{eq:dhs}\\
r_{\mathrm{out}}&=&\frac{r_{\mathrm{V}}}{(1-f_{\mathrm{v}})^{1/3}},\ r_{\mathrm{in}}=\frac{f_{\mathrm{v}}^{1/3}r_{\mathrm{V}}}{(1-f_{\mathrm{v}})^{1/3}},
\end{eqnarray}
where $f_{\mathrm{v}}$ is the fraction of vacuum of the hollow sphere, $n(f_{\mathrm{v}})df_{\mathrm{v}}$ is a number of hollow spheres in a range $[f_{\mathrm{v}},f_{\mathrm{v}}+df_{\mathrm{v}}]$, $A(r_{\mathrm{in}},r_{\mathrm{out}})$ represents an optical quantity of the hollow sphere with inner and outer radii of $r_{\mathrm{in}}$ and $r_{\mathrm{out}}$, and $r_{\mathrm{V}}$ is the volume equivalent radius. The distribution function is assumed to have \citep{M05}
\begin{eqnarray}
n(f_{\mathrm{v}})=
\left\{
\begin{array}{ll}
1/f_\mathrm{max} & (0\leq f_{\mathrm{v}}< f_\mathrm{max}),\\
0 & (f_{\mathrm{v}}\ge f_\mathrm{max}),\\
\end{array}
\right.
\end{eqnarray}
where $f_{\mathrm{max}}$ is the free parameter of the DHS method, and it is sometimes referred to as the irregularity parameter. In this paper, we adopt $f_{\mathrm{max}}=0.8$ \citep{M16}.
Since each monomer of dust aggregates has the homogeneous silicate composition in our dust model, we use astronomical silicate for the shell composition of the hollow spheres. 
Optical properties of a hollow sphere are obtained by using the code \texttt{dmilay\_f95.f} \citep{T81}, which is included in OpacityTool of the DIANA standard opacities \citep{Woitke:2016aa}.

In Figure \ref{fig:kappa1}, we compare the opacities obtained by DHS to those obtained by TMM.

First of all, we discuss the absorption opacities (top panels in Figure \ref{fig:kappa1}).
Unlike the Mie theory and EMT, DHS can show large absorption opacities at the Rayleigh domain ($kR_g<1$), and this is one of the important properties of the DHS method.
The form factor distribution of the DHS particle model is somewhat similar to that of fractal dust aggregates \citep{Min:2006aa}, and hence, DHS succeeds to produce an enhancement of the absorption opacity at the Rayleigh domain (see Section \ref{sec:excess} for more detail discussion). 

Outside the Rayleigh domain ($kR_g>1$), it is found that DHS fails to reproduce the TMM results.
It is also shown that the opacities obtained by the DHS method is similar to those of the Mie theory.
As we mentioned, the absorption opacities of BCCA and BPCA are characterized by the monomer's property when $\Delta\phi<1$. 
Because the DHS method does not include the information of the monomer radius in the formulation, it is not surprising that DHS fails to reproduce the TMM results. 

Secondly, we explain the scattering opacities of BCCA and BPCA models (middle panels). Outside the Rayleigh domain ($kR_g>1$), DHS does not reproduce the scattering opacities of TMM.
Similar to the case of the absorption opacity, the scattering opacities obtained by DHS are very similar to those obtained by the Mie theory. Because in the DHS method, the scattering opacity is averaged over the distribution of hollow spheres, interference in the scattering opacities are smeared out; hence, the DHS results show more smooth wavelength dependence compared to those obtained by the Mie theory.

Thirdly, we discuss the single scattering albedo.
Since DHS does not reproduce both absorption and scattering properties for BCCA and BPCA outside the Rayleigh domain, the albedo at this wavelength domain deviates from the TMM results. 

As a result, we conclude that DHS is not able to reproduce both absorption and scattering opacities of BCCA and BPCA at $kR_g\gtrsim 1$. \citet{M16} showed that DHS seems to capture the opacities of compact dust aggregates. This difference is presumably due to the fact that both BCCA and BPCA are porous enough (porosity $\gtrsim 85\%$) so that optical properties of monomers tend to govern the optical properties of the dust aggregates. Since DHS does not include the information of the monomer in the calculations, it fails to reproduce the optical properties of dust aggregates.
If the dust aggregates are compact enough so that optical properties are governed by bulk properties of dust aggregates rather than the monomer, DHS is considered to be a good approximation as shown in \citet{M16}.

\subsection{Asymmetry parameter and effective albedo} \label{sec:compare2}

\begin{figure*}
\begin{center}
\includegraphics[height=6.0cm,keepaspectratio]{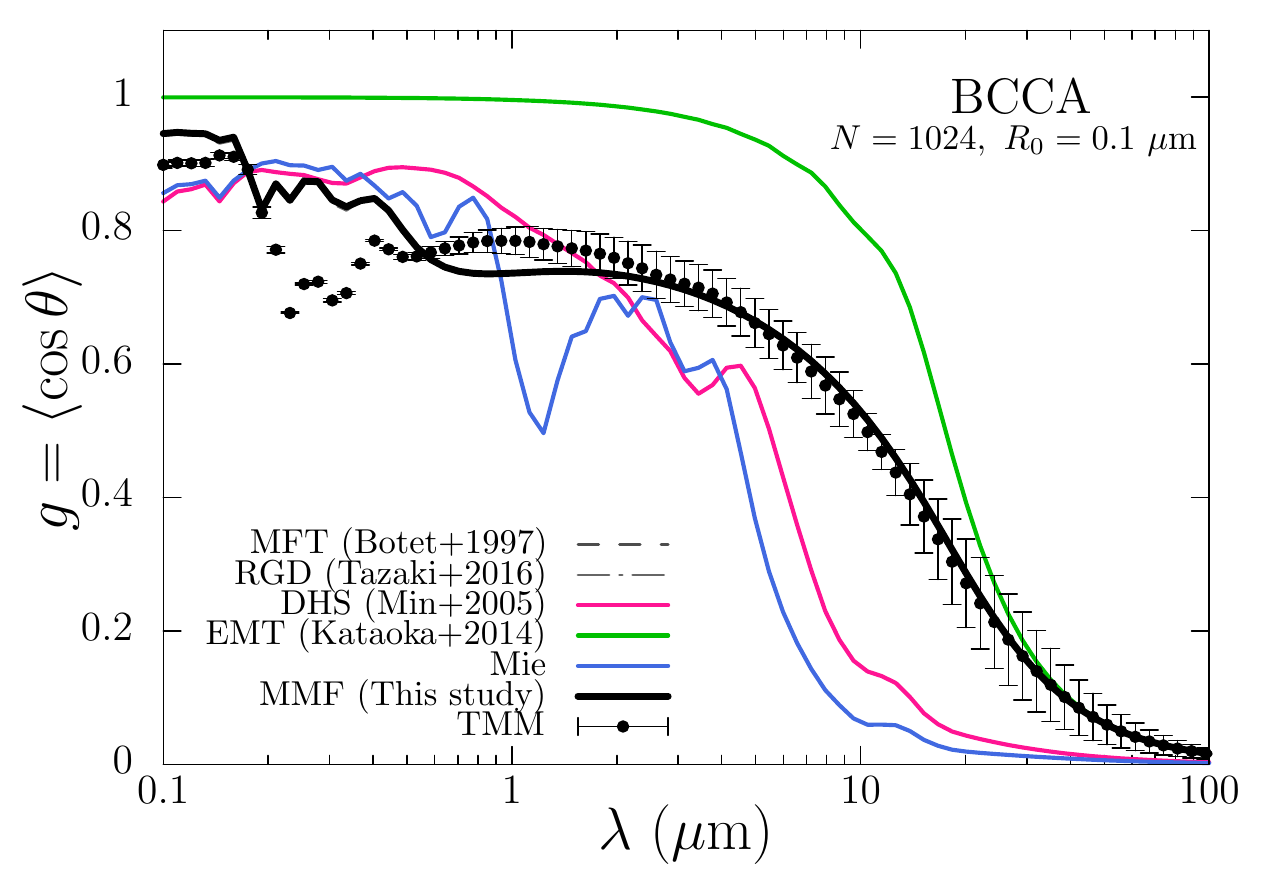}
\includegraphics[height=6.0cm,keepaspectratio]{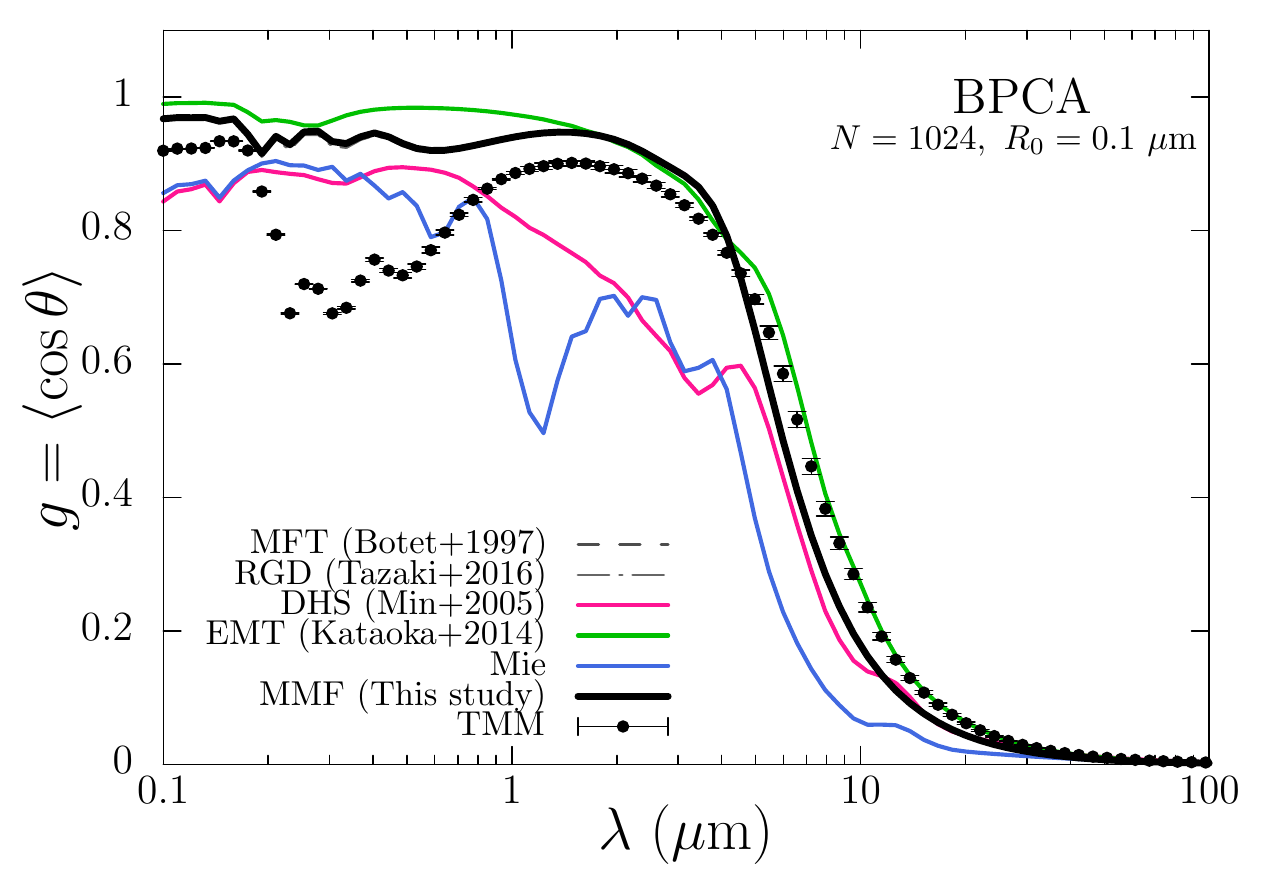}
\includegraphics[height=6.0cm,keepaspectratio]{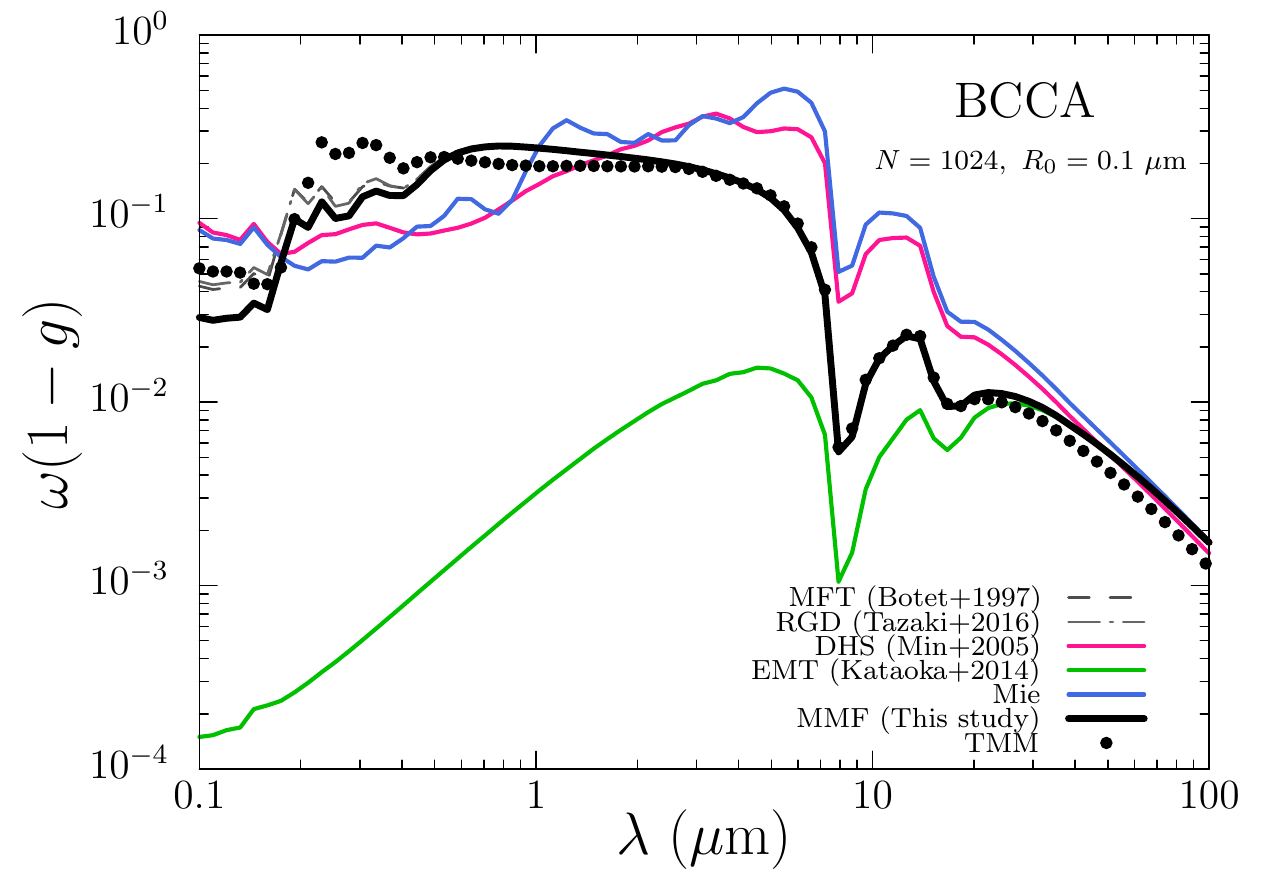}
\includegraphics[height=6.0cm,keepaspectratio]{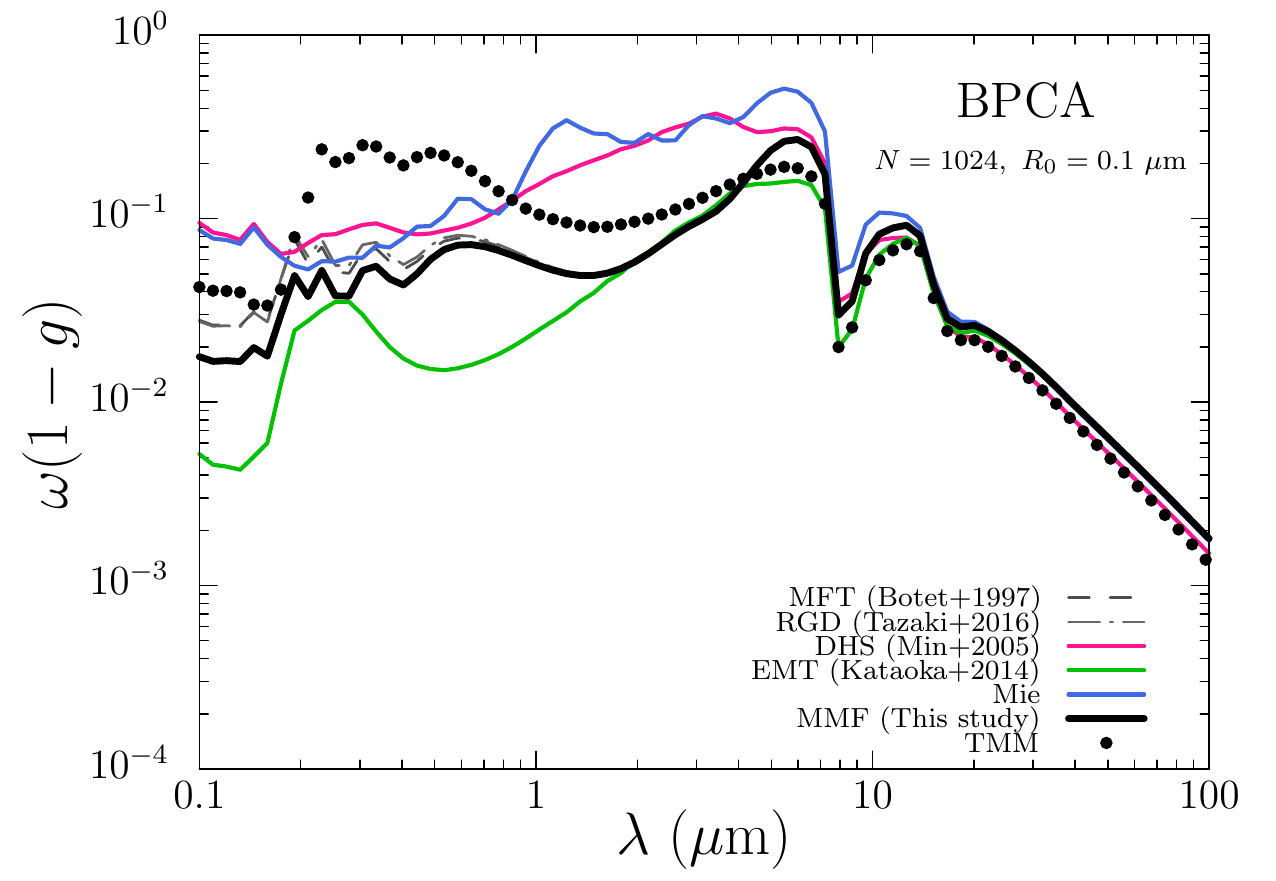}
\caption{Same as Figure \ref{fig:kappa1}, but for the asymmetry parameter and the effective albedo.}
\label{fig:comp2}
\end{center}
\end{figure*}
Figure \ref{fig:comp2} shows the asymmetry parameter $g$ and the effective albedo $(1-g)\omega$.
The $g$ parameter represents the degree of forward scattering, e.g., $g=0$ means isotropic scattering and $g=1$ means perfect forward scattering.
When the aggregate is in the Rayleigh domain ($kR_g<1$), scattering becomes isotropic, and hence, $g=0$. Once the wavelength becomes shorter than the size of dust aggregates ($kR_g>1$), forward scattering happens, and hence, $g$ increases.

First of all, we discuss the asymmetry parameter of BCCA.
As long as the single scattering assumption is validated ($\Delta\phi<1$), the angular dependence of scattering can be understood in terms of the static structure factor \citep[see][]{T16}. Since, in the single scattering domain, EMT tends to significantly underestimate the side- and back-scattering intensity due to destructive interference \citep{T16}, the $g$ value quickly approaches to unity as decreasing the wavelength. However, as shown in \citet{T16}, BCCA does not show an efficient destruction of scattered waves. As a result, the wavelength dependence of the $g$ value is much softer than the EMT results. Since MMF, MFT, and the RGD theory take the realistic static structure factor into account, they can reproduce the wavelength dependence of $g$. At the multiple scattering region ($\Delta\phi>1$), the scattering phase function does not necessary to obey the static structure factor. Therefore, MMF, MFT, and the RGD theory may not be a predictable theory for the phase function in the multiple scattering domain, although opacities can be correctly calculated even in the multiple scattering domain. Next, we discuss the asymmetry parameter of BPCA. In this case, the single scattering assumption can be readily violated, and then the $g$ parameter obtained by MMF, MFT, and the RGD theory is less accurate than the case of BCCA. Since DHS does not take the static structure factor into account, it cannot reproduce the phase function as well as the asymmetry parameter at the single scattering domain. As a result, DHS fails to reproduce the wavelength dependence of $g$ for both BPCA and BCCA outside the Rayleigh domain.

Next, we discuss the effective albedo of the aggregate particles. 
Since forward scattering is hardly observable for most cases, it is useful to define an albedo without including the contribution of forward scattering.
In this paper, we adopt a simple definition of the effective albedo, $(1-g)\omega$.
Suppose particles show perfect forward scattering ($g=1$), then incident light seems to just pass through the particles without changing the direction; hence, the observer (supposed to be at $\theta\ne 0$) does not see any {\it scattered} light. In this case, the effective albedo $(1-g)\omega$ gives zero. If particles show isotropic scattering ($g=0$), the effective albedo gives the single scattering albedo $\omega$. 

At the bottom panels of Figure \ref{fig:comp2}, we show the effective albedo of the BCCA and BPCA models.
In case of BCCA, the effective albedo obtained by EMT is two orders of magnitude smaller than the TMM results at near-infrared wavelength. Again, this is due to the destructive interference of EMT, and then, EMT give rise to the extremely strong forward scattering which makes the effective albedo significantly low. In addition, color of the effective albedo obtained by EMT is reddish. This is due to the fact that forward scattering becomes strong for shorter wavelengths. As a result, if EMT is applied to BCCA, scattering of BCCA shows very faint and reddish color. However, the TMM results show that this behavior is not true. The effective albedo of BCCA is not so faint as expected from EMT and shows gray color instead of reddish color.

This result have an interesting implication to the scattered light observations of protoplanetary disks. \citet{Mulders:2013aa} pointed out the color and brightness problem of scattered light images of protoplanetary disks and suggested that this problem might be solved by considering the presence of large fluffy dust aggregates (porosity $>99\%$). However, our results imply that large fluffy dust aggregates, like BCCA, do not show faint and reddish color in the effective albedo. We will not go into the details of this topic here because it is beyond the scope of this paper. However, this topic will be discussed in more detail in our forthcoming paper.

\section{Absorption property of the aggregate in the Rayleigh domain}  \label{sec:excess}
 
We investigate the absorption opacity of dust aggregates in the Rayleigh domain.
As mentioned in Sections \ref{sec:syserror} and \ref{sec:approx}, the absorption opacity at the Rayleigh domain obtained by TMM is larger than that obtained by the Mie theory, the RGD theory, MFT, MMF, and EMT (MG-Mie).
In this section, we study a physical process that is responsible for the enhancement of the absorption opacity.

It is worth mentioning that the enhanced absorption of dust aggregates would be important to explain the large millimeter-wave opacity inferred from observations of protoplanetary disks.

\subsection{Origin of the enhanced absorption}
\subsubsection{Qualitative argument}
The enhanced absorption of dust aggregates in the Rayleigh domain has been reported by a number of authors \citep{Wright:1987aa,bazell90,kozasa92, Mackowski:1995aa, Henning:1995aa, Stognienko:1995aa, Mackowski:2006aa, Kohler:2011aa}\footnote{The enhanced absorption reported by \citet{bazell90} is slightly larger than the other studies. As pointed out in \citet{Rouleau:1993aa}, this is because \citet{bazell90} adopted the lax convergence criterion in their DDA calculations.}. 

We can qualitatively understand the origin of the enhanced absorption of dust aggregates by a simple argument presented in \citet{Henning:1995aa}.
\citet{Henning:1995aa} considered that spheres form a straight chain elongated parallel to the incident electric field and approximated this sphere cluster by a cylinder. For the case of an isolated single sphere, the internal electric field is weakened by the depolarization effect of surface charges, which appear as a consequence of electrostatic polarization of the particle. However, for the case of a sufficiently long cylinder, depolarization can be negligible for incident light polarized along the elongation \citep[see e.g., p. 147 of ][]{B83}; hence, the internal electric field can be as strong as the external field. As a result, strong absorption happens. Since dust aggregates may contain a chain-like pattern being almost orientated to the incident field, this effect is expected to remain in dust aggregates. 
Thus, in general, the absorption opacity in the Rayleigh domain depends on the particle shape through the depolarization factor. 

\subsubsection{Enhanced absorption of a two-sphere cluster}
\citet{Henning:1995aa} approximated an elongated sphere cluster as a cylinder. 
Because the particle shape sensitively affects its absorption property, we need to directly solve the internal field of a sphere cluster \citep{Mackowski:1995aa, Mackowski:2006aa} for quantitative estimate of the enhanced absorption.

First of all, we study the enhanced absorption of a very simple system, that is, a two-sphere cluster.
We perform TMM calculations of a two-sphere cluster with a fixed orientation using \textsc{mstm-v3.0} code \footnote{The code and more information is available on \url{http://www.eng.auburn.edu/~dmckwski/scatcodes/}}. Each spherical particle has the radius $R_0=0.1\ \mu$m, and the incident wavelength is set as $\lambda=1\ $mm. Therefore, both of a whole cluster and each spherical monomer are in the Rayleigh domain. The refractive index is set as astronomical silicate at $\lambda=1\ $mm, and hence, $m=3.434+0.049i$. 

\begin{figure*}
\begin{center}
\includegraphics[height=5.5cm,keepaspectratio]{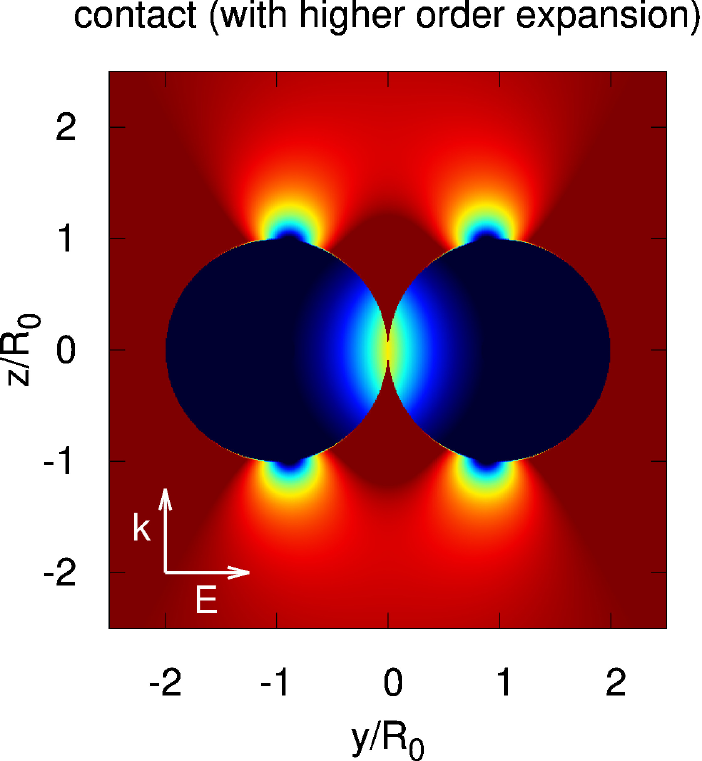}
\includegraphics[height=5.5cm,keepaspectratio]{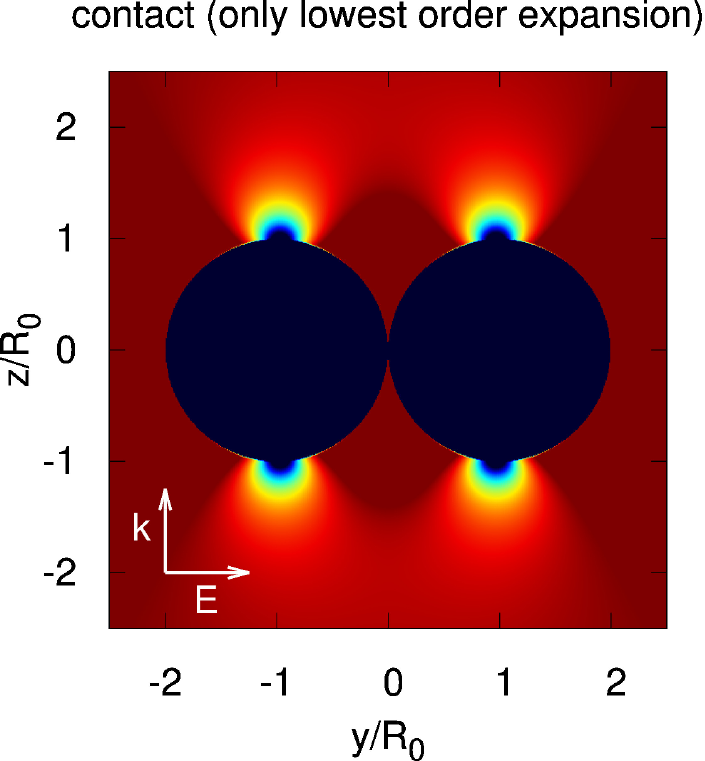}
\includegraphics[height=5.5cm,keepaspectratio]{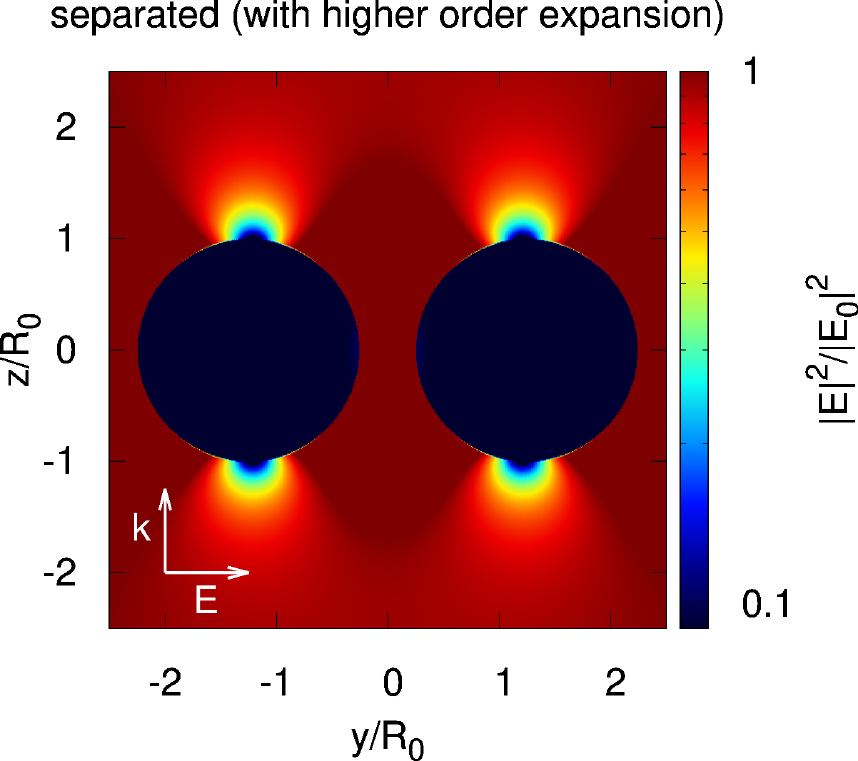}
\caption{
The electric field strength map for a two-sphere cluster at the cross section
including centers of two spheres. The electric field is normalized by
the electric field strength of the incident light $E_0$.
Both vertical and horizontal axes are normalized by the monomer radius $R_0$. A two-sphere cluster is elongated along $y$ direction (horizontal axis), while the incident light polarized along $y$ direction is propagating along $z$ direction (vertical axis). (left) Two spheres are in contact, and for each sphere, higher order of harmonic expansion is included ($N_O=10$), (middle) two spheres are in contact, and for each sphere, only dipole term is included ($N_O=1$), and (right) two spheres are separated by the center-to-center distance of $1.5R_0$, and higher order expansions is included.}
\label{fig:emap}
\end{center}
\end{figure*}

Figure \ref{fig:emap} show the electric field strength map at the cross section including the centers of two spheres. In the map, both internal and external electric fields of the cluster are shown. 
It can be seen from Figure \ref{fig:emap} that the strong electric field appears in the vicinity of a contact point of two spheres.
When the harmonic expansion of each particle is truncated at the dipole term ($N_O=1$), the electric field in the vicinity of the contact point becomes weak. In addition, when two spheres are separated, the internal electric field becomes weak. Therefore, the enhanced absorption is due to the proximity effect of the neighboring monomer, and higher order terms of the harmonic expansion are important for this effect.

In Table \ref{tab:excessbi}, the absorption cross section of a two-sphere cluster, which is normalized by the individual sum of the monomer's absorption cross section, $NC_{\mathrm{abs}}^0$,  are summarized. 
Note that the normalization of the Table \ref{tab:excessbi} values, $NC_{\mathrm{abs}}^0$, equals to the absorption cross section of the cluster obtained by the RGD theory, MFT, MMF, EMT (MG-Mie) and the Mie theory with a volume equivalent sphere. Note that not only the RGD theory, but also MFT, MMF, EMT (MG-Mie) and the Mie theory with a volume equivalent sphere show the absorption cross section $NC_{\mathrm{abs}}^0$ (see Figure \ref{fig:kappa1}).
Thus, values given in Table \ref{tab:excessbi} may be regarded as the deviation of these methods from TMM at the Rayleigh domain with $m=3.434+0.049i$. From Table \ref{tab:excessbi}, we can see that the calculation included only dipole expansion underestimates the absorption opacity of the cluster by around 16\% for astronomical silicate at $\lambda=1$ mm for contacting spheres. Once two monomers are separated, interaction between monomers becomes weak, and hence, the absorption cross section of two spheres approaches to the individual sum of each absorption cross section.

This results clearly demonstrate that the importance of higher order expansion for the monomer in the Rayleigh domain. Even if the monomer particle radius is much smaller than the wavelength, higher order terms of harmonic expansion can largely contribute to the absorption opacity when the refractive index is large. Hence, magnitude of the enhanced absorption reported by studies \citep[e.g.,][]{bazell90,kozasa92, Stognienko:1995aa} should be regarded as the lower limit, because in these studies, each spherical monomer is replaced by a single dipole moment. 
Although \citet{Stognienko:1995aa} adopt dipole approximation for each monomer (in their DDA model), it should be mentioned that they have also examined multipole approximation for each monomer where they considered at most 4th order multipole expansion (in their DMM model).

\begin{table}
\caption{Cross sections for a two-sphere cluster in the Rayleigh domain}
\label{tab:r}
\centering
\begin{tabular}{lccc}
\hline
 & \multicolumn{3}{c}{$C_{\mathrm{abs}}/NC_{\mathrm{abs}}^0$\footnote{A two-sphere cluster is elongated along the $\hat{\mathrm{y}}$ direction. }} \\
\cline{2-4}
model & unpolarized & $\mathbf{E} || \hat{\mathbf{y}}$ & $\mathbf{E} \perp \hat{\mathbf{y}}$ \\
\hline  \hline
contact ($N_O=10$)  &  $1.345$  &  $1.835$ & 0.856\\
contact ($N_O=1$)  &  $1.188$  &  $1.546$ & 0.830\\
separated ($3R_0/2$)\footnote{This separation distance is center-to-center distance of a two-sphere cluster.}  & 1.084   & 1.256  & 0.911\\
separated ($2R_0$)  &  1.038 & 1.131 & 0.945\\
\hline
\end{tabular}
\label{tab:excessbi}
\end{table}

\subsection{Enhanced absorption of dust aggregates}
We study the enhanced absorption of sphere clusters with more complicated structure and discuss how the number of monomers, the aggregate structure, and the composition affect it.

We consider two types of dust aggregates,  BCCA and BPCA, and number of monomers is varied from 1 (single monomer) to 1024. Although \citet{Mackowski:2006aa} has already studied the enhanced absorption for the cluster-cluster aggregation model, we focus on more compact dust aggregates as well as BCCA.

We adopt the same numerical method and set up as Section \ref{sec:setup}, except for the refractive index.
Refractive index of the monomer is taken from astronomical silicate or amorphous carbon at $\lambda=1$ mm. 
Optical constants of amorphous carbon is taken from the BE model of \citet{Zubko:1996aa}, and it has $m=12.67+4.597i$. The material density of amorphous carbon is set as $1.8$ g cm$^{-3}$. Note that Because we adopt $N_O=10$ for amorphous carbon, the result for amorphous carbon might be underestimated, and hence it should be regarded as the lower limit.

\begin{figure*}
\begin{center}
\includegraphics[height=5.5cm,keepaspectratio]{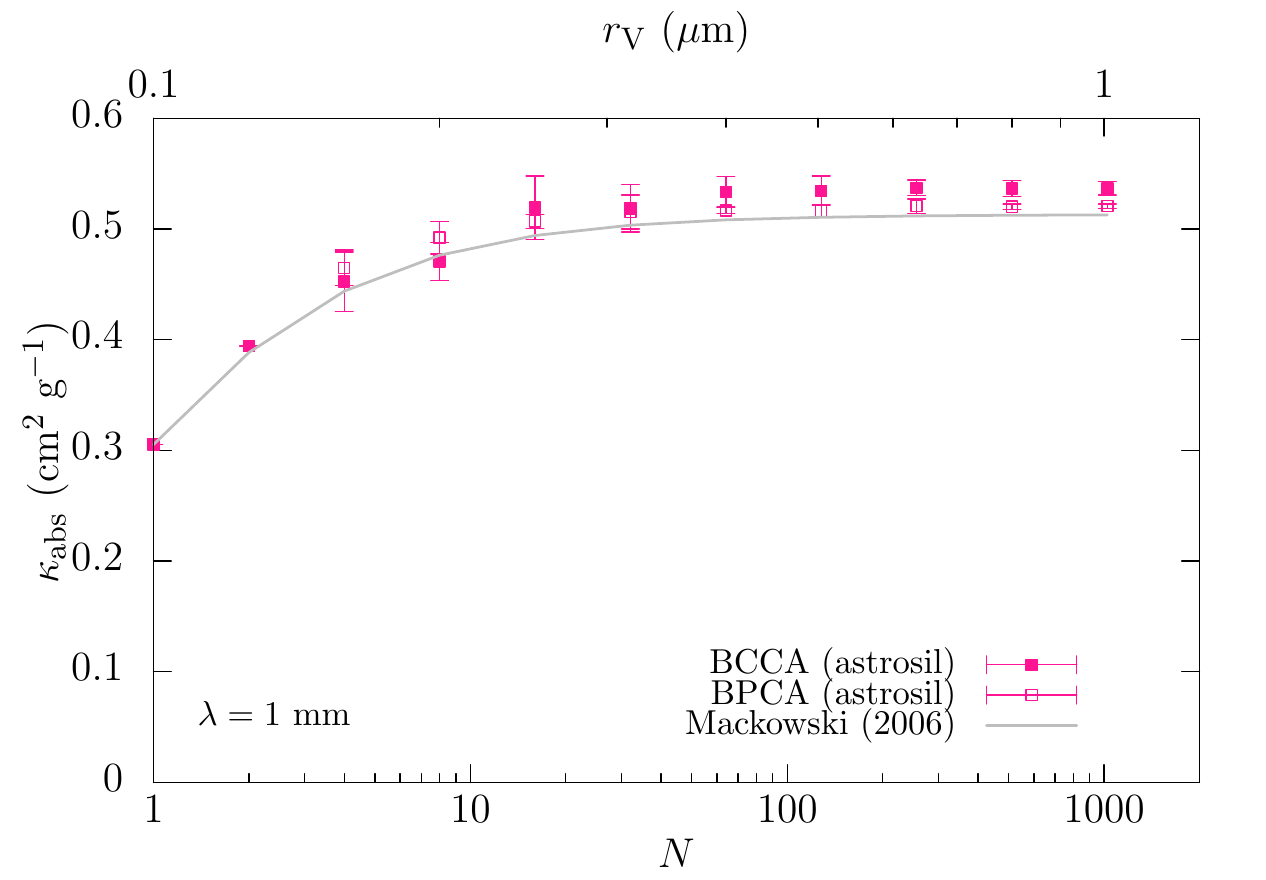}
\includegraphics[height=5.5cm,keepaspectratio]{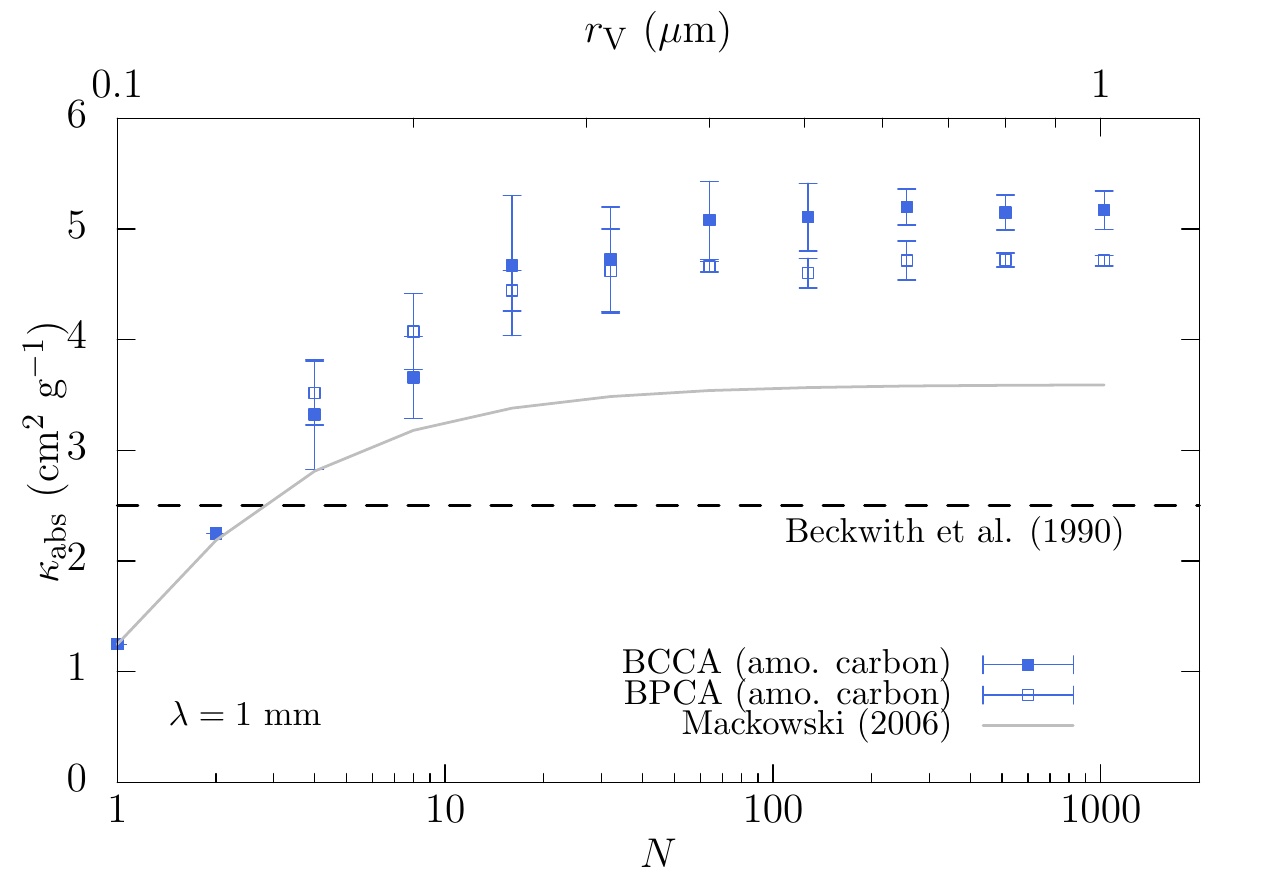}
\caption{
Absorption opacities of BPCA and BCCA for astronomical silicate (left)
and amorphous carbon (right) in the Rayleigh domain.
Horizontal dashed line in the right panel represents the opacity value used in \citet{Beckwith:1990aa}.
Refractive indices at $\lambda=1$ mm are $m=3.434+0.049i$ and $12.67+4.597i$ for astronomical silicate and amorphous carbon, respectively.}
\label{fig:excess}
\end{center}
\end{figure*}

Figure \ref{fig:excess} shows the absorption opacity of dust aggregates. As increasing the number of monomers, the absorption opacity also increases. However, once the number of monomers exceeds around $N\sim10^{2}$, the absorption opacity does not depend on the number of monomers.
Surprisingly, the absorption opacity of BPCA is very close to that of BCCA, although their overall shapes are significantly different. 
In addition, the enhancement of the absorption opacity is large for the dust aggregates with large refractive index. Therefore, the dust aggregates with amorphous carbon show large excess in the absorption opacity compared to the case of astronomical silicate.
In this case of amorphous carbon, even if the volume equivalent aggregate radius is $\simeq 1\ \mu$m, the absorption opacity of the aggregate is significantly enhanced. 

At $N=1024$, dust aggregates with amorphous carbon show $\kappa_{\mathrm{abs}}\simeq 5$ cm$^2$ per gram of dust. 
This value is much higher than the commonly used opacity value in millimeter-wave observations of protoplanetary disks \citep{Beckwith:1990aa}, whereas the opacity for $N=1$ is smaller than its value. 
Our results demonstrate that the large opacity value adopted in \citet{Beckwith:1990aa} can be readily achieved by considering the proximity effect of dust aggregates in the Rayleigh domain. Therefore, the millimeter-wave opacity value presented in \citet{Beckwith:1990aa} is possibly explained not only mm-sized compact grains \citep{DAlessio01}, but also small dust aggregates. 

\begin{figure}
\begin{center}
\includegraphics[height=6cm,keepaspectratio]{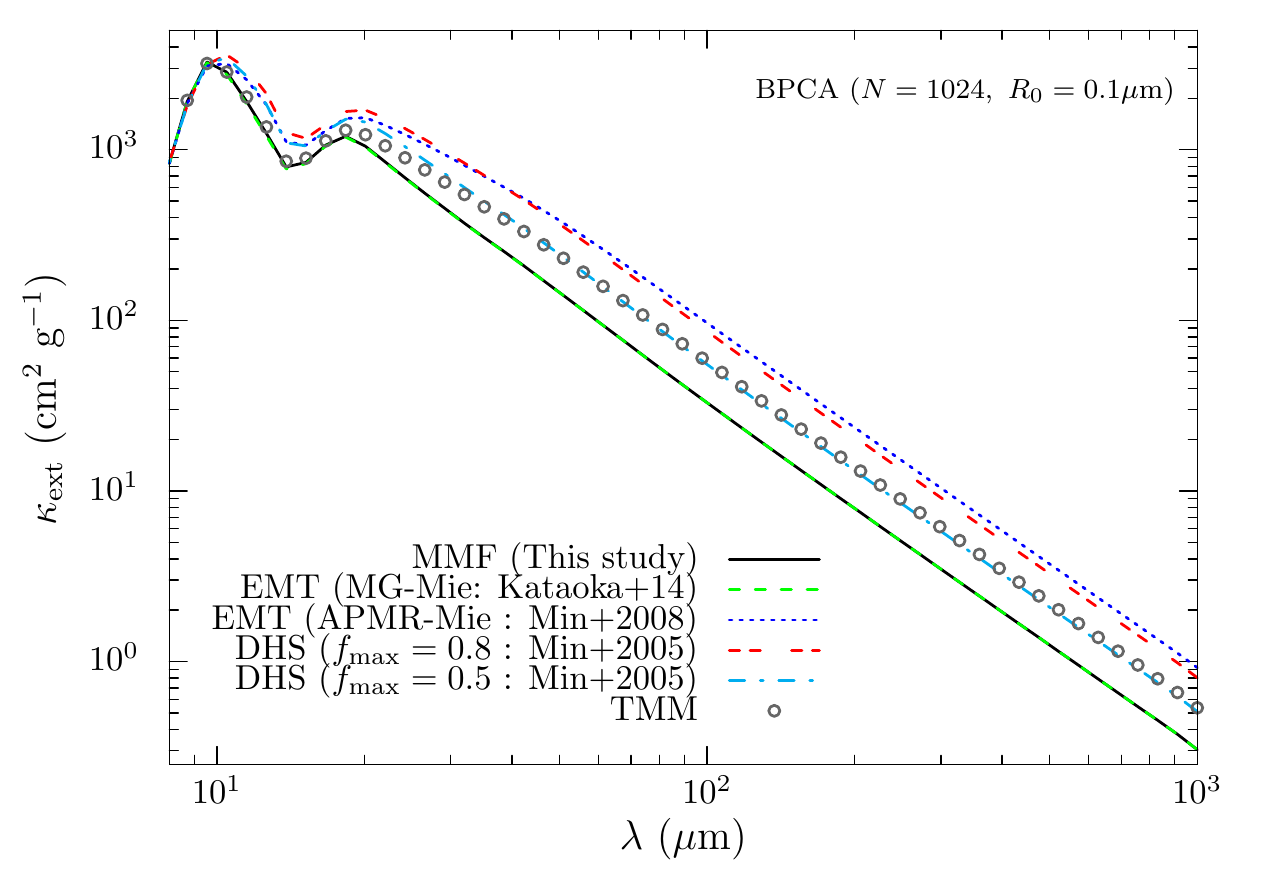}
\caption{Extinction opacity of BPCA with $N=1024$ and $R_0=0.1\ \mu$m obtained by EMT and DHS. Circles indicate the results obtained by TMM. Various lines are obtained by MMF (This study), EMT with MG-Mie and the APMR method \citep{Min:2008aa}, and the DHS method with $f_{\mathrm{max}}=0.8$ and 0.5 \citep{M05}.}
\label{fig:apmr}
\end{center}
\end{figure}

\subsection{Can we approximate the enhanced absorption?}
We compare the enhanced absorption of approximate methods; EMT and DHS. In addition, we consider two types of mixing rules in the EMT calculations: MG-Mie and APMR-Mie.

\subsubsection{MMF, MFT, and the RGD theory}
According to Equation (\ref{eq:absrgd}), MMF, MFT and the RGD theory predict that the absorption opacity of dust aggregates is the same as that of the monomer in the Rayleigh limit. Figure \ref{fig:apmr} shows that the absorption opacity obtained by MMF is smaller than that obtained by TMM. This is because the connection effect is not included in MMF, MFT, and the RGD theory. 

\subsubsection{EMT: APMR-Mie}
\citet{Min:2008aa} proposed a mixing rule so-called aggregate polarizability mixing rule (APMR).
APMR is the same as the Maxwell-Garnett mixing rule except for the polarizability of inclusions.
In the Maxwell-Garnett mixing rule, polarizability of the inclusion is determined by assuming the spherical inclusions.
In APMR, the polarizability is given by the CDE model so that the effect of irregularly shaped monomer can be taken into account. It is worth noting that the CDE model has similar distribution of form factor with fractal dust aggregates of spherical monomers \citep{Min:2006aa}, and hence EMT (APMR-Mie) might be used to approximate the enhanced absorption of dust aggregates.

We can write an explicit form of the effective dielectric function of APMR:
\begin{equation}
\epsilon_{\rm eff}^{\mathrm{APMR}}=\frac{3(\epsilon_{\rm i}-1)+4f[\epsilon_{\rm i}\ln{\epsilon_{\rm i}}-(\epsilon_{\rm i}-1)]}
{3(\epsilon_{\rm i}-1)-2f[\epsilon_{\rm i}\ln{\epsilon_{\rm i}}-(\epsilon_{\rm i}-1)]}, \label{eq:APMReff} 
\end{equation}
where 
$\epsilon_{\rm i}$ is the dielectric function of inclusions, and we have assumed that the dielectric function of matrix $\epsilon_{\rm m}=1$.
Figure \ref{fig:apmr} show the extinction opacity obtained by EMT with Maxwell-Garnett mixing rule and APMR. Note that we use the characteristic radius $R_c$ in order to define the filling factor the of dust aggregate, whereas \citet{Min:2008aa} used the radius of gyration. 

Since the Maxwell-Garnett rule mixing assumes the individual sum of polarizability of spherical monomers as the total polarizability of dust aggregates, EMT (MG-Mie) does not include the proximity effect. Therefore, it coincide with the MMF result in the Rayleigh limit, and then the absorption opacity obtained by EMT (MG-Mie) is underestimated compared to the TMM result in the Rayleigh limit.

The APMR method also assumes individual sum of CDE polarizability; however, EMT (APMR-Mie) can show the enhanced absorption. This is because polarizability of the ellipsoidal particle is larger than that of sphere with the same volume. 
The EMT (APMR) seems to overestimate the absorption opacity; however, it should be keep in mind that our TMM results should be regarded as the lower limit (see Section \ref{sec:shortsum}).

\subsubsection{DHS}
In Figure \ref{fig:apmr}, we also plot the opacity obtained by DHS with $f_{\mathrm{max}}=0.8$. In DHS, the absorption opacity at the Rayleigh domain can be tuned by the irregularity parameter $f_{\mathrm{max}}$. When $f_{\mathrm{max}}=0.5$, DHS gives rise to the similar absorption opacity to that obtained by TMM. 

Figure \ref{fig:apmr} shows that when $f_{\mathrm{max}}=0.8$, the opacities of EMT (APMR-Mie) and DHS at the Rayleigh domain are quite similar. It is known that there is a similarity between CDE and DHS. This similarity can be reduced to the similarity of polarizability between a hollow sphere and an ellipsoid in the Rayleigh limit \citep{M03, Min:2006aa}. In the case of a randomly orientated ellipsoid, polarizability is increased when the axis ratio is increased.
In the case of a hollow sphere, polarizability in the Rayleigh domain can be given by 
\begin{equation}
\alpha=3V_\mathrm{m}\frac{(\epsilon-1)(2\epsilon+1)}{(\epsilon+2)(2\epsilon+1)-2f_{\mathrm{v}}(\epsilon-1)^2}, \label{eq:hs}
\end{equation}
where $V_\mathrm{m}$ is the material volume \citep{B83}. $f_\mathrm{v}=0$ gives the polarizability of the homogeneous sphere.
{\it When the material volume is fixed}, increasing $f_\mathrm{v}$ results in an increase of the polarizability $\alpha$. 
In the analogy of ellipsoids, increasing $f_{\mathrm{v}}$ is similar to decreasing the depolarization factor \citep[see e.g., Chapter 5 of][for the definition of the depolarization factor]{B83}. Therefore, averaging hollow spheres over a distribution of various $f_{\mathrm{v}}$ is similar to the average of ellipsoids with various axis ratio.

\subsection{Effects not considered in this study} \label{sec:shortsum}
Our TMM calculations are based on spherical monomers with a point connection. However, realistic monomers are connected via the non-zero contact area. \citet{Kohler:2011aa} studied the enhanced absorption for various connection states, e.g., the presence of neck structure, of the sphere cluster and concluded that the contact area can largely affect the quantitative estimate of the enhanced absorption when the refractive index is large. 

Another effect that is not considered in our TMM calculations is non-sphericity of the monomer. 
Since non-sphericity of the monomer, such as elongation and surface roughness, increases the absorption cross section of the single monomer, it is expected that introducing non-sphericity of monomers further increases the absorption opacity of the dust aggregates.

Therefore, the absorption opacities in the Rayleigh domain obtained in our TMM calculations should be regarded as the lower limit. 
We expect that when the non-sphericity of the monomer and a neck structure between monomers are taken into account, the absorption opacity becomes larger than that obtained by our TMM calculations, and it may be close to that estimated by EMT (APMR-Mie) or DHS with $f_{\mathrm{max}}=0.8$.

\section{Conclusion} \label{sec:conclusion}
By using various numerical methods, we have calculated the opacities
for absorption and scattering of two types of fractal dust aggregates, BCCA and BPCA, which have fractal dimensions of 2 and 3, respectively.
The opacities obtained by the TMM method, a rigorous numerical method, were compared to those obtained by approximate methods, the RGD theory, MFT, EMT and DHS. 
The results in our comparison can be summarized as follows.
\begin{enumerate}
\item The extinction opacity of BCCA and BPCA obtained by TMM are well reproduced by using MFT for all parameters we have tested (see Section \ref{sec:wavel}). On the other hand, the RGD-theory breaks down when multiple scattering is important ($\Delta\phi>1$).
\item MFT fails to reproduce absorption and scattering opacities, once multiple scattering starts to dominate the scattering ($\Delta\phi>1$).
The absorption opacity at the multiple scattering domain is underestimated by at most 94\% and 87\% for the BPCA and BCCA models, respectively. We showed that this is due to the fact that the mean field assumption gives rise to the incorrect estimate of the single scattering albedo at the multiple scattering domain (see also Equations (\ref{eq:mftsca} and \ref{eq:mftabs})).
\item We propose a simple modification of MFT, which was referred to as MMF. We showed that MMF can reproduce the absorption and scattering opacities at the multiple scattering domain (see Section \ref{sec:mmf}), as well as the asymmetry parameter (see Section \ref{sec:compare2}).
MFT enables to compute opacities of fractal dust aggregates rapidly and accurately.
The advantage of MMF is that information of monomer and fractal dimension (aggregates structure) are retained in the formulation; therefore, we can study how monomers and 
aggregate structures affect observational properties of astrophysical environments, such as protoplanetary disks and planetary atmosphere.
\item EMT tends to underestimate the absorption opacity when the monomer size parameter becomes large for the BCCA and BPCA models. Although it seems to be able to reproduce scattering opacity of BPCA, EMT significantly underestimates
the scattering opacity of BCCA when $\lambda<2\pi R_g$. EMT shows prominent forward scattering (see Paper I), and hence, EMT underestimates the effective albedo of BCCA by an order of magnitude (see Section \ref{sec:emttmm}). 
\item DHS does not reproduce the absorption and scattering opacities of BCCA and BPCA at the wavelengths shorter than the size of dust aggregates (see Section \ref{sec:dhstmm}). This is presumably due to the fact that these aggregates are porous enough (porosity $\gtrsim 85\%$) so that the monomers tends to govern the optical properties, whereas the DHS method does not include the information of monomer in the calculation.
\item 
Coagulation of dust particles leads to an enhancement of the absorption opacity due to the proximity effect. The enhanced absorption opacity of dust aggregates in the Rayleigh domain obtained by the TMM calculations can be roughly reproduced DHS with $f_{\mathrm{max}}=0.5$. However, it is important to keep in mind that our TMM estimates of the absorption opacity in the Rayleigh domain should be regarded as the lower limit (see Section \ref{sec:excess}).
\end{enumerate}

\acknowledgments
We sincerely thank the referee for a thorough and careful reading of the manuscript.
R.T. would like to thank Daniel Mackowski and Yasuhiko Okada for the availability of the T-Matrix code with the QMC method. R. T. also thank Bruce Draine for making the DDSCAT code public. 
R.T. thank Robert Botet and Pascal Rannou for kindful advices about the mean field theory and its implementation and also thank Akimasa Kataoka and Satoshi Okuzumi for useful discussion. R.T. was supported by a Research Fellowship for Young Scientists from the Japan Society for the Promotion of Science (JSPS) (17J02411).
\appendix

\section{Benchmark test of mean field theory calculations} \label{sec:app1}
Coefficients $\bar{A}_{1,n}^{1,\nu}, \bar{B}_{1,n}^{1,\nu}$ can by given by \citep{B97}
\begin{eqnarray}
\bar{A}_{1,n}^{1,\nu}&=&\frac{2\nu+1}{n(n+1)\nu(\nu+1)}\sum_{p=|n-\nu|}^{n+\nu}[n(n+1)+\nu(\nu+1)
-p(p+1)]a(\nu,n,p)s_p(kR_g) \\
\bar{B}_{1,n}^{1,\nu}&=&2\frac{2\nu+1}{n(n+1)\nu(\nu+1)}\sum_{p=|n-\nu|}^{n+\nu} b(\nu,n,p)s_p(kR_g).
\end{eqnarray}
$a(\nu,n,p)$ and $b(\nu,n,p)$ are defined by 
\begin{eqnarray}
a(\nu,n,p)&=&\frac{2p+1}{2}\int_{-1}^{1} P_{\nu}^{1}(x)P_{n}^{1}(x)P_{p}(x)dx \label{eq:anunp}\\
b(\nu,n,p)&=&\frac{2p+1}{2}\int_{-1}^{1} P_{\nu}^{1}(x)P_{n}^{1}(x)\frac{dP_{p}(x)}{dx}dx \label{eq:bnunp}
\end{eqnarray}
where $P_n^m(x)$ is the associated Legendre function.
The structure of the dust aggregate is specified by
\begin{equation}
s_p(kR_g)=\frac{\pi^2}{k^3}\int_0^{\infty} uJ_{p+1/2}(u)H_{p+1/2}^{(1)}(u)g(u/k)du \label{eq:sp}
\end{equation}
where $J_{p+1/2}(u)$ is the Bessel function, and $H_{p+1/2}^{(1)}(u)$ is the Hankel function.

When $p$ has the same parity with $n+\nu$, the coefficient $B$ always vanishes because integrand of Equation (\ref{eq:bnunp}) becomes odd function, whereas the coefficient $A$ becomes not-zero because integrand of Equation (\ref{eq:anunp}) becomes even function. On the other hand, when $p$ does not have the same parity with $n+\nu$, the coefficient $A$ vanishes, whereas the coefficient $B$ becomes non-zero. Above parity argument is useful to speed up the computation of the mean field values.

We numerically integrate $a(\nu,n,p), b(\nu,n,p)$, and $s_p(kR_g)$ instead of using recurrence relations given in Appendix of \citet{B97}. In Table \ref{tab:dn12}, mean field values of first 5 orders for reaction-limited cluster-cluster aggregates (RCCA) model are shown. Parameters used to obtain the mean field values in Table \ref{tab:dn12} are as follows. RCCA has fractal dimension of $d_f=2.0$ and fractal prefactor of $k_0=0.825$ \citep{B95}. We adopt $\lambda=0.8\  \mu$m, $R_0=0.5\ \mu$m, $N=64$, and complex refractive index as $m=1.4+0.0001i$. Although our results shows $\bar{d}_{1,1}^{(1)}=(0.352,0.245)$, whereas \citet{B97} shows $\bar{d}_{1,1}^{(1)}=(0.996,0.041)$ with the same parameters.
We also compare our results to those obtained by the mean field code implemented by Pascal Rannou, and two results agrees well. As a result, we speculate that this mismatch is due to typographical errors in \citet{B97}, noting that it is impossible to check because the original code is not available any more (Robert Botet and Pascal Rannou, in private communication). 

\begin{table}
\caption{Coefficients of the mean field for the first 5 orders of the RCCA model}
\label{tab:r}
\centering
\begin{tabular}{lcc}
\hline
$n$  & $\bar{d}_{1,n}^{(1)}$ & $\bar{d}_{1,n}^{(2)}$ \\
\hline  \hline
1 &    $0.352+0.245i$  &   $0.396+0.124i$\\
2  &  $0.419+0.119i$   &  $0.408+0.179i$\\
3  &   $0.388-0.039i$ &    $0.448+0.093i$\\
4   &  $0.123-0.111i$   & $0.067-0.079i$\\
5    & $0.014-0.024i$  &  $0.005-0.009i$\\
\hline
\end{tabular}
\label{tab:dn12}
\end{table}

\section{Test of different cut off models of two-points correlation function} \label{sec:app2}
We investigate how the choice of cut-off models given in Equation (\ref{eq:comodel}) affects the value of the extinction opacity. We calculate the extinction opacities of BCCA and BPCA. The monomer radius and the number of monomers are set as $R_0=0.1\ \mu$m and $N=1024$, respectively. The refractive index is assumed to be astronomical silicate.

\begin{figure}[t]
\begin{center}
\includegraphics[height=6.0cm,keepaspectratio]{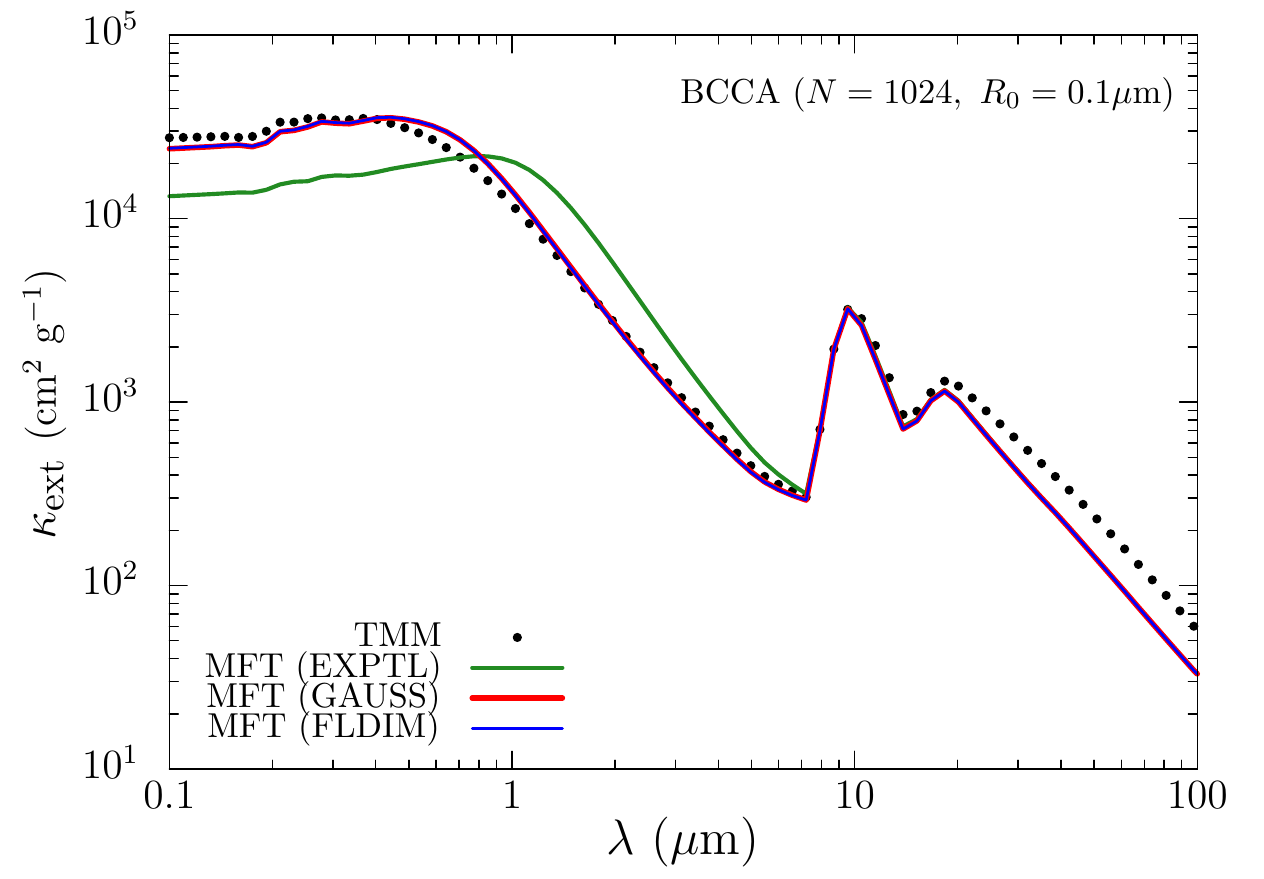}
\includegraphics[height=6.0cm,keepaspectratio]{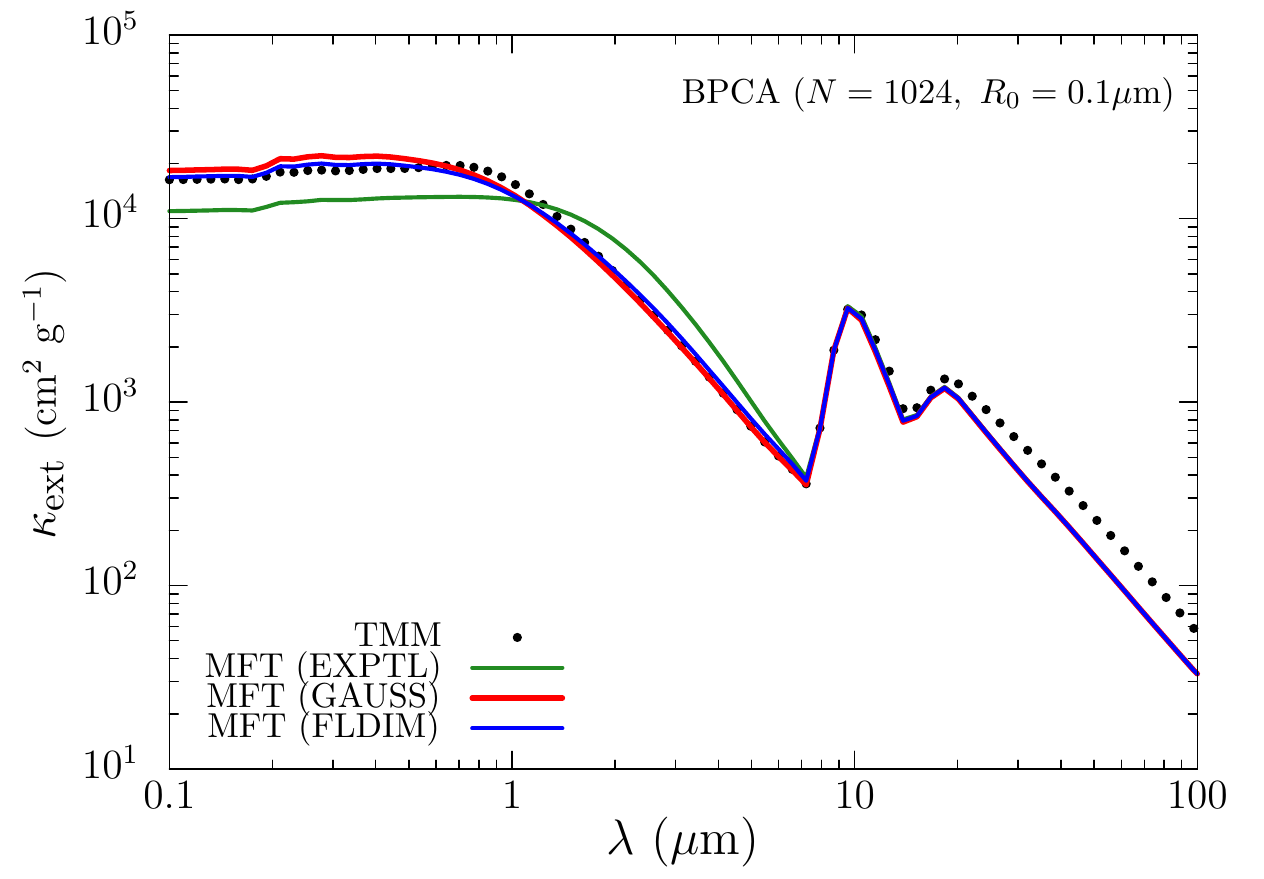}
\caption{Extinction opacity for BCCA (left panel) and BPCA (right panel). Dots represent the TMM results, while lines indicate the MFT results. Green, red, and blue lines represent the exponential cut-off model, the Gaussian cut-off model, and the fractal dimension cut-off model, respectively (see Equation (\ref{eq:comodel})). }
\label{fig:correlation}
\end{center}
\end{figure}

Figure \ref{fig:correlation} shows the extinction opacities of the BCCA and BPCA models obtained by MFT. 
In the case of the BCCA model, the FLDIM and GAUSS models show almost similar results, whereas the EXPTL model fails to reproduce the extinction opacity obtained by TMM. For the BPCA model, both the GAUSS and FLDIM models can produce the extinction cross section to an accuracy less than 20\%. 
Therefore, we recommend to use the GAUSS or FLDIM cut-off model.
One advantage to use the GAUSS model is that integration of the static structure factor (see Equation (\ref{eq:sq})) can be done analytically \citep[see][]{T16}, and hence, the fast computation is possible compared to the FLDIM model. On the other hand, an advantage of using the FLDIM model is that it is slightly more accurate than the GAUSS model.

\section{Behavior of Opacities calculated by MFT} \label{sec:BP}
The behavior of MFT can be captured by considering a scalar wave scattering theory \citep{B86}. Suppose the scalar wave incident on $j$th monomer can be denoted by $\psi_j=c_j\exp[i\bm{k}_i\cdot\bm{r}_j]$, where $\bm{k}_i$ is the wave vector of the (externally) incident light, $\bm{r}_j$ is the position vector of $j$th monomer, and $c_j$ represents the effect of multiple scattering, e.g., $c_j=1$ means single scattering. The mean field assumption implies that $c_j$ must be the same for every monomer; hence, we write $c=c_j$=constant. 
According to \citet{B86}, the self-consistent solution to $c$ becomes
\begin{eqnarray}
c=[1-(N-1)\tau Q]^{-1}, \label{eq:cvalue}
\end{eqnarray}
where $\tau$ is related to the scattering amplitude of a single monomer via $\tau/k$, and $Q$ is a quantity that depends on the aggregate structure and the wavelength \citep[for more detail, see][]{B86}. 
For large $N$, the $c$ value is governed by the quantity $N\tau Q$. \citet{B86} showed that $|N\tau Q|$ has following dependences:
\begin{eqnarray}
|N\tau Q| &\propto& (kR_0)^{3-d_f}\ (d_f<2), \label{eq:NtQ1}\\
|N\tau Q| &\propto& kR_0N^{1-2/d_f}\ (d_f>2). \label{eq:NtQ2}
\end{eqnarray}
It follows from Equations (\ref{eq:NtQ1},\ref{eq:NtQ2}) that $|N\tau Q|$ is irrelevant to $N$ for $d_f<2$, whereas $|N\tau Q|$ increases with increasing $N$ for $d_f>2$. 

When $|N\tau Q|\ll1$, $|c|\simeq1$, and hence, single scattering is the dominant process. In this case, MFT coincides with the RGD theory. As increasing $|N\tau Q|$, the effect of multiple scattering increases. When $|N\tau Q|$ is slightly less than unity (still multiple scattering is sub-dominant), we have $|c|>1$; hence multiple scattering strengthens the incident field. It can be seen from Figure \ref{fig:f1} that at the wavelengths where the phase shift is slightly smaller than unity, the extinction opacities of MFT is slightly larger than those of the RGD theory. Once $|N\tau Q|$ exceeds unity, $|c|<1$. This means that multiple scattering has a role in reducing the net incident field. Because multiple scattering weakens the incident field, the extinction opacities obtained by the MFT become smaller than those of the RGD theory.

According to \citet{B86}, scattering and absorption cross sections of MFT can be rewritten by 
\begin{eqnarray}
C_{\mathrm{sca}}^{(\mathrm{MFT})}&=&|c|^2C_{\mathrm{sca}}^{(\mathrm{RGD})}, \label{eq:mftsca}\\ 
C_{\mathrm{abs}}^{(\mathrm{MFT})}&=&|c|^2C_{\mathrm{abs}}^{(\mathrm{RGD})}=|c|^2NC_{\mathrm{abs,mono}}. \label{eq:mftabs}
\end{eqnarray}
It follows from Equations (\ref{eq:mftsca} and \ref{eq:mftabs}) that single scattering albedo, $\omega=C_{\mathrm{sca}}^{(\mathrm{MFT})}/C_{\mathrm{ext}}^{(\mathrm{MFT})}$, is irrelevant to multiple scattering factor $|c|$. In other words, the albedo obtained by MFT is the same as that of the RGD theory. Since the RGD theory is a single scattering theory, it is incapable of reproducing optical properties at the multiple scattering domain ($\Delta\phi>1$). Therefore, the scattering and absorption opacities obtained by MFT becomes incorrect at the multiple scattering domain.


\begin{thebibliography}{}

\bibitem[{{Bazell} \& {Dwek}(1990)}]{bazell90}
{Bazell}, D., \& {Dwek}, E. 1990, \apj, 360, 142

\bibitem[Beckwith et al.(1990)]{Beckwith:1990aa} Beckwith, S.~V.~W., Sargent, A.~I., Chini, R.~S., \& Guesten, R.\ 1990, \aj, 99, 924 

\bibitem[{{Berry} \& {Percival}(1986)}]{B86}
{Berry}, M.~V., \& {Percival}, I.~C. 1986, Optica Acta, 33, 577

\bibitem[{{Blum} \& {Wurm}(2008)}]{Blum:2008aa}
{Blum}, J., \& {Wurm}, G. 2008, \araa, 46, 21

\bibitem[{{Bohren} \& {Huffman}(1983)}]{B83}
{Bohren}, C.~F., \& {Huffman}, D.~R. 1983, {Absorption and scattering of light
  by small particles}

\bibitem[{{Botet} {et~al.}(1995){Botet}, {Rannou}, \& {Cabane}}]{B95}
{Botet}, R., {Rannou}, P., \& {Cabane}, M. 1995, Journal of Physics A
  Mathematical General, 28, 297

\bibitem[{{Botet} {et~al.}(1997){Botet}, {Rannou}, \& {Cabane}}]{B97}
---. 1997, \ao, 36, 8791

\bibitem[{{Cuzzi} {et~al.}(2014){Cuzzi}, {Estrada}, \& {Davis}}]{cuzzi14}
{Cuzzi}, J.~N., {Estrada}, P.~R., \& {Davis}, S.~S. 2014, \apjs, 210, 21

\bibitem[D'Alessio et al.(2001)]{DAlessio01} D'Alessio, P., Calvet, N., \& Hartmann, L.\ 2001, \apj, 553, 321 

\bibitem[{{Draine} \& {Flatau}(1994)}]{draine94}
{Draine}, B.~T., \& {Flatau}, P.~J. 1994, \josaa, 11, 1491

\bibitem[{{Draine} \& {Lee}(1984)}]{D84}
{Draine}, B.~T., \& {Lee}, H.~M. 1984, \apj, 285, 89

\bibitem[{{Henning} {et~al.}(1995){Henning}, {Michel}, \&
  {Stognienko}}]{Henning:1995aa}
{Henning}, T., {Michel}, B., \& {Stognienko}, R. 1995, \planss, 43, 1333

\bibitem[Henning \& Stognienko(1996)]{Henning96} Henning, T., \& Stognienko, R.\ 1996, \aap, 311, 291 

\bibitem[{{Kataoka} {et~al.}(2014){Kataoka}, {Okuzumi}, {Tanaka}, \&
  {Nomura}}]{K14}
{Kataoka}, A., {Okuzumi}, S., {Tanaka}, H., \& {Nomura}, H. 2014, \aap, 568,
  A42

\bibitem[{{Kimura} {et~al.}(2016){Kimura}, {Kolokolova}, {Li}, \&
  {Lebreton}}]{Kimura:2016aa}
{Kimura}, H., {Kolokolova}, L., {Li}, A., \& {Lebreton}, J. 2016, ArXiv
  e-prints, arXiv:1603.03123

\bibitem[{{K{\"o}hler} {et~al.}(2011){K{\"o}hler}, {Guillet}, \&
  {Jones}}]{Kohler:2011aa}
{K{\"o}hler}, M., {Guillet}, V., \& {Jones}, A. 2011, \aap, 528, A96


\bibitem[{{Kozasa} {et~al.}(1992){Kozasa}, {Blum}, \& {Mukai}}]{kozasa92}
{Kozasa}, T., {Blum}, J., \& {Mukai}, T. 1992, \aap, 263, 423

\bibitem[{{Kozasa} {et~al.}(1993){Kozasa}, {Blum}, {Okamoto}, \&
  {Mukai}}]{kozasa93}
{Kozasa}, T., {Blum}, J., {Okamoto}, H., \& {Mukai}, T. 1993, \aap, 276, 278

\bibitem[{{Laor} \& {Draine}(1993)}]{L93}
{Laor}, A., \& {Draine}, B.~T. 1993, \apj, 402, 441

\bibitem[Mackowski(1994)]{Mackowski:1994aa} Mackowski, D.~W.\ 1994, Journal of the Optical Society of America A, 11, 2851 

\bibitem[{{Mackowski}(1995)}]{Mackowski:1995aa}
{Mackowski}, D.~W. 1995, \ao, 34, 3535

\bibitem[{{Mackowski}(2006)}]{Mackowski:2006aa}
---. 2006, \jqsrt, 100, 237

\bibitem[{{Mackowski} \& {Mishchenko}(1996)}]{M96}
{Mackowski}, D.~W., \& {Mishchenko}, M.~I. 1996, Journal of the Optical Society
  of America A, 13, 2266

\bibitem[{{Marley} {et~al.}(2013){Marley}, {Ackerman}, {Cuzzi}, \&
  {Kitzmann}}]{Marley:2013aa}
{Marley}, M.~S., {Ackerman}, A.~S., {Cuzzi}, J.~N., \& {Kitzmann}, D. 2013,
  {Clouds and Hazes in Exoplanet Atmospheres}, ed. S.~J. {Mackwell}, A.~A.
  {Simon-Miller}, J.~W. {Harder}, \& M.~A. {Bullock}, 367--391

\bibitem[{{Maxwell Garnett}(1904)}]{MG09}
{Maxwell Garnett}, J.~C. 1904, Royal Society of London Philosophical
  Transactions Series A, 203, 385

\bibitem[{{Min} {et~al.}(2012){Min}, {Canovas}, {Mulders}, \&
  {Keller}}]{Min:2012aa}
{Min}, M., {Canovas}, H., {Mulders}, G.~D., \& {Keller}, C.~U. 2012, \aap, 537,
  A75

\bibitem[{{Min} {et~al.}(2003){Min}, {Hovenier}, \& {de Koter}}]{M03}
{Min}, M., {Hovenier}, J.~W., \& {de Koter}, A. 2003, \aap, 404, 35

\bibitem[{{Min} {et~al.}(2005){Min}, {Hovenier}, \& {de Koter}}]{M05}
---. 2005, \aap, 432, 909

\bibitem[{{Min} {et~al.}(2006){Min}, {Hovenier}, {Dominik}, {de Koter}, \&
  {Yurkin}}]{Min:2006aa}
{Min}, M., {Hovenier}, J.~W., {Dominik}, C., {de Koter}, A., \& {Yurkin}, M.~A.
  2006, \jqsrt, 97, 161

\bibitem[{{Min} {et~al.}(2008){Min}, {Hovenier}, {Waters}, \& {de
  Koter}}]{Min:2008aa}
{Min}, M., {Hovenier}, J.~W., {Waters}, L.~B.~F.~M., \& {de Koter}, A. 2008,
  \aap, 489, 135

\bibitem[{{Min} {et~al.}(2016){Min}, {Rab}, {Woitke}, {Dominik}, \&
  {M{\'e}nard}}]{M16}
{Min}, M., {Rab}, C., {Woitke}, P., {Dominik}, C., \& {M{\'e}nard}, F. 2016,
  \aap, 585, A13

\bibitem[{{Minato} {et~al.}(2006){Minato}, {K{\"o}hler}, {Kimura}, {Mann}, \&
  {Yamamoto}}]{M06}
{Minato}, T., {K{\"o}hler}, M., {Kimura}, H., {Mann}, I., \& {Yamamoto}, T.
  2006, \aap, 452, 701

\bibitem[{{Mishchenko} {et~al.}(1996){Mishchenko}, {Travis}, \&
  {Mackowski}}]{mishchenko96}
{Mishchenko}, M.~I., {Travis}, L.~D., \& {Mackowski}, D.~W. 1996, \jqsrt, 55,
  535

\bibitem[{{Mishchenko} {et~al.}(2016){Mishchenko}, {Dlugach}, {Yurkin}, {Bi},
  {Cairns}, {Liu}, {Panetta}, {Travis}, {Yang}, \&
  {Zakharova}}]{Mishchenko:2016aa}
{Mishchenko}, M.~I., {Dlugach}, J.~M., {Yurkin}, M.~A., {et~al.} 2016,
  \physrep, 632, 1


\bibitem[{{Mukai} {et~al.}(1992){Mukai}, {Ishimoto}, {Kozasa}, {Blum}, \&
  {Greenberg}}]{Mukai:1992aa}
{Mukai}, T., {Ishimoto}, H., {Kozasa}, T., {Blum}, J., \& {Greenberg}, J.~M.
  1992, \aap, 262, 315
  
\bibitem[Mulders et al.(2013)]{Mulders:2013aa} Mulders, G.~D., Min, M., Dominik, C., Debes, J.~H., \& Schneider, G.\ 2013, \aap, 549, A112

\bibitem[{{Okada}(2008)}]{O08}
{Okada}, Y. 2008, \jqsrt, 109, 1719

\bibitem[Okuzumi et al.(2012)]{Okuzumi:2012aa} Okuzumi, S., Tanaka, H., Kobayashi, H., \& Wada, K.\ 2012, \apj, 752, 106

\bibitem[{{Ormel} {et~al.}(2011){Ormel}, {Min}, {Tielens}, {Dominik}, \&
  {Paszun}}]{Ormel:2011aa}
{Ormel}, C.~W., {Min}, M., {Tielens}, A.~G.~G.~M., {Dominik}, C., \& {Paszun},
  D. 2011, \aap, 532, A43

\bibitem[{{Ormel} {et~al.}(2009){Ormel}, {Paszun}, {Dominik}, \&
  {Tielens}}]{Ormel:2009aa}
{Ormel}, C.~W., {Paszun}, D., {Dominik}, C., \& {Tielens}, A.~G.~G.~M. 2009,
  \aap, 502, 845

\bibitem[Ossenkopf(1991)]{Ossenkopf91} Ossenkopf, V.\ 1991, \aap, 251, 210 

\bibitem[{{Rouleau} \& {Martin}(1993)}]{Rouleau:1993aa}
{Rouleau}, F., \& {Martin}, P.~G. 1993, \apj, 416, 707

\bibitem[Purcell \& Pennypacker(1973)]{purcell73} Purcell, E.~M., \& Pennypacker, C.~R.\ 1973, \apj, 186, 705

\bibitem[{{Shen} {et~al.}(2008){Shen}, {Draine}, \& {Johnson}}]{shen08}
{Shen}, Y., {Draine}, B.~T., \& {Johnson}, E.~T. 2008, \apj, 689, 260

\bibitem[{{Shen} {et~al.}(2009){Shen}, {Draine}, \& {Johnson}}]{shen09}
---. 2009, \apj, 696, 2126

\bibitem[Silsbee \& Draine(2016)]{Silsbee:2016aa} Silsbee, K., \& Draine, B.~T.\ 2016, \apj, 818, 133

\bibitem[{Sorensen(2001)}]{S01}
Sorensen, C.~M. 2001, Aerosol Science and Technology, 35, 648

\bibitem[{{Steinacker} {et~al.}(2010){Steinacker}, {Pagani}, {Bacmann}, \&
  {Guieu}}]{Steinacker:2010aa}
{Steinacker}, J., {Pagani}, L., {Bacmann}, A., \& {Guieu}, S. 2010, \aap, 511,
  A9

\bibitem[{{Stognienko} {et~al.}(1995){Stognienko}, {Henning}, \&
  {Ossenkopf}}]{Stognienko:1995aa}
{Stognienko}, R., {Henning}, T., \& {Ossenkopf}, V. 1995, \aap, 296, 797

\bibitem[Paper I()]{T16}
{Tazaki}, R., {Tanaka}, H., {Okuzumi}, S., {Kataoka}, A., \& {Nomura}, H. 2016,
  \apj, 823, 70 (Paper I)

\bibitem[{{Toon} \& {Ackerman}(1981)}]{T81}
{Toon}, O.~B., \& {Ackerman}, T.~P. 1981, \ao, 20, 3657

\bibitem[Voshchinnikov et al.(2005)]{vosh05} Voshchinnikov, N.~V., Il'in, V.~B., \& Henning, T.\ 2005, \aap, 429, 371 

\bibitem[{{Voshchinnikov} {et~al.}(2007){Voshchinnikov}, {Videen}, \&
  {Henning}}]{vosh07}
{Voshchinnikov}, N.~V., {Videen}, G., \& {Henning}, T. 2007, \ao, 46, 4065

\bibitem[{{Woitke} {et~al.}(2016){Woitke}, {Min}, {Pinte}, {Thi}, {Kamp},
  {Rab}, {Anthonioz}, {Antonellini}, {Baldovin-Saavedra}, {Carmona}, {Dominik},
  {Dionatos}, {Greaves}, {G{\"u}del}, {Ilee}, {Liebhart}, {M{\'e}nard},
  {Rigon}, {Waters}, {Aresu}, {Meijerink}, \& {Spaans}}]{Woitke:2016aa}
{Woitke}, P., {Min}, M., {Pinte}, C., {et~al.} 2016, \aap, 586, A103

\bibitem[{{Wright}(1987)}]{Wright:1987aa}
{Wright}, E.~L. 1987, \apj, 320, 818

\bibitem[Zubko et al.(1996)]{Zubko:1996aa} Zubko, V.~G., Mennella, V., Colangeli, L., \& Bussoletti, E.\ 1996, \mnras, 282, 1321 

\end{thebibliography}
\end{document}